\documentclass[12pt]{article}

\usepackage{amssymb, amsmath, a4}
\usepackage{cite}
\def\be{\begin{equation}}
\def\ee{\end{equation}}
\def\bea{\begin{eqnarray}}
\def\eea{\end{eqnarray}}
\def\beb{\begin{eqnarray*}}
\def\eeb{\end{eqnarray*}}
\allowdisplaybreaks[4] 
\makeatletter
\def\fmslash{\@ifnextchar[{\fmsl@sh}{\fmsl@sh[0mu]}}
\def\fmsl@sh[#1]#2{%
  \mathchoice
    {\@fmsl@sh\stackrelsplaystyle{#1}{#2}}%
    {\@fmsl@sh\textstyle{#1}{#2}}%
    {\@fmsl@sh\scriptstyle{#1}{#2}}%
    {\@fmsl@sh\scriptscriptstyle{#1}{#2}}}
\def\@fmsl@sh#1#2#3{\m@th\ooalign{$\hfil#1\mkern#2/\hfil$\crcr$#1#3$}}
\makeatother

\begin{document}


\thispagestyle{empty}
\begin{titlepage}

\begin{flushright}
hep-th/0703169\\
Preprint ESI 1900 (2007)\\
UWThPh-2007-4\\
 
\end{flushright}

\vspace{0.3cm}
\boldmath
\begin{center}
  {\large {\bf Induced Gauge Theory on a Noncommutative Space }}
\end{center}
\unboldmath
\vspace{0.8cm}
\begin{center}
                                                                                                                                         
{\bf
Harald~Grosse\footnote{Email: harald.grosse@univie.ac.at} and
Michael~Wohlgenannt\footnote{Supported by the Erwin Schr\"odinger International Institute 
for Mathematical Physics (ESI), Vienna and 
by the Austrian Science Fund, project P18657-N16. Email: michael.wohlgenannt@univie.ac.at}
}
                                                                                                                                         
\end{center}
\vskip 1em
\begin{center}
Universit\"at Wien, Fakult\"at f\"ur Physik, \\
  Boltzmanngasse 5, A-1090 Wien, Austria\\
\end{center}
\vskip 1em


\vfill
\begin{abstract}
\noindent
We consider a scalar $\phi^4$ theory on canonically deformed Euclidean space in $4$
dimensions with an additional oscillator potential. This model is known to be renormalisable.
An exterior gauge field is coupled in a 
gauge invariant manner to the scalar field. We extract the dynamics for the
gauge field from the divergent terms of the 1-loop effective action using a matrix basis
and propose an action for the noncommutative gauge theory, which is a candidate for a renormalisable
model.
\end{abstract}

\end{titlepage}


\section{\label{introduction}Introduction}

Feynman rules for Quantum Field Theory over noncommutative spaces lead for planar diagrams to 
the standard renormalisation problem, for non-planar ones an additional problem running under the
name of infrared / ultraviolet mixing shows up.

In a previous work \cite{Grosse:2003nw,Grosse:2004yu}, the structure of divergences is studied carefully, and it is realised that
the expanded model with four marginal operators leads to a renormalisable theory,
\bea
S_0 =  \int d^4 x &\left( \frac{1}{2} \phi \star 
[\tilde x_\nu, \,  [\tilde x^\nu, \phi]_\star]_\star 
+ \frac{\Omega^2}{2} 
\phi \star \{ \tilde x^\nu , \{ \tilde x_\nu ,\phi\}_\star \}_\star 
\right.
\nonumber
\\
& \left. + \frac{\mu^2}{2} \phi \star \phi 
+ \frac{\lambda}{4!} \phi\star \phi  \star \phi \star \phi\right)(x)\;.
\label{action1}
\eea
This model fulfills the Langmann-Szabo duality \cite{Langmann:2002cc} which motivates the added term.
There are various proofs of renormalisability available \cite{Rivasseau:2005bh,Gurau:2005gd}. 
Similar results for fermion models have 
also been obtained by the Paris group \cite{Vignes-Tourneret:2006nb}.
We restrict ourselves to the canonical Euclidean space with constant commutation relations
\be
[x^\mu \stackrel{\star}{,} x^\nu] = i \theta^{\mu\nu},
\ee
where $\theta^{ij}=-\theta^{ji}\in \mathbb{R}$,
and the $\star$-product is given by the Weyl-Moyal product
\be
f\star g \, (x) = e^{i\theta^{\mu\nu}\frac{\partial}{\partial x^\mu}\frac{\partial}{\partial y^\nu}}
f(x)g(y)\big|_{y\to x}.
\ee
The differential calculus is generated by 
$$
\partial_\mu f = -i[\tilde x_\mu, f]_\star\,.
$$

In order to obtain the action for a gauge theory, which hopefully is renormalisable, we extract the divergent
terms of the heat kernel expansion. Such a procedure leads in the 
commutative case to a renormalisable gauge field action.
We introduce the local, unitary gauge group $\mathcal G$ under which the scalar field $\phi$ transforms covariantly
like
\be
\label{i.1}
\phi \mapsto u^* \star \phi \star u, \,\, u\in \mathcal G.
\ee

The approach employed here makes use of two basic ideas. 
First, it is well known that the $\star$-multiplication of a coordinate - and also of a function, of 
course - with a field is not a covariant process. The product $x^\mu \star \phi$ will not transform
covariantly,
$$
x^\mu \star \phi \nrightarrow u^* \star x^\mu \star \phi \star u\;.
$$
Functions of the coordinates are not effected by the gauge group. The matter field $\phi$ is taken to be
an element of a module \cite{Jurco:2000fs}. The introduction of covariant coordinates 
\be
\tilde X_\nu=\tilde x_\nu + A_\nu  
\ee
finds a remedy to this situation \cite{Madore:2000en}. The gauge field $A_\mu$ and hence the
covariant coordinates transform in the
following way:
\bea
\label{i.2}
A_\mu & \mapsto & \mathrm{i} u^* \star \partial_\mu u  + u^* \star A_\mu
\star u \,, \\
\nonumber  
\tilde X_\mu & \mapsto & u^* \star \tilde X_\mu \star u
\; .
\eea
Using covariant coordinates we can construct an action invariant under gauge transformations. This
action defines the model we are going to study in this article:
\bea
S & = & \int d^4 x \left( 
\frac{1}{2} \phi \star [\tilde X_\nu,\, [\tilde X^\nu,\, \phi]_\star ]_\star 
+ \frac{\Omega^2}{2} 
\phi \star \{\tilde X^\nu , \{ \tilde X_\nu ,\phi\}_\star \}_\star 
\right.
\nonumber
\\
&&
+ \left. \frac{\mu^2}{2} \phi \star \phi 
+ \frac{\lambda}{4!} \phi\star \phi  \star \phi \star \phi\right)(x)\;.
\label{action}
\eea

Secondly, we apply the heat kernel formalism. The gauge field $A_\mu$ is an external, classical gauge
field coupled to $\phi$. In the following sections, we will explicitly 
calculate the divergent terms of the one-loop effective action. In the classical case, the divergent
terms determine the dynamics of the gauge field
\cite{Chamseddine:1996zu,Langmann:2001cv,Vassilevich:2003xt}.
There have already been attempts to generalise this approach to the non-commutative realm; for
non-commutative $\phi^4$ theory see \cite{Gayral:2004cs,Gayral:2004cu}.
First steps towards gauge kinetic models have been done in
\cite{Vassilevich:2003yz,Gayral:2004ww,Vassilevich:2005vk}. However, the results there are not completely comparable,
since we have modified the free action and expand around $-\nabla^2 + \Omega^2 \tilde x^2$ rather than $-\nabla^2$.

A few days ago, A.~de~Goursac, J.-Chr.~Wallet and R.~Wulkenhaar \cite{deGoursac:2007gq} published a 
paper where they computed the effective action for a similar model in coordinate
space. They have evaluated relevant Feynman diagrams 
and obtained the same results as presented here.

We note that the general formalism developed by A.~Connes and A.~Chamseddine \cite{Connes:2006qj} cannot be applied here, since in
our case a tadpole contribution shows up, which is supposed to vanish in their work.

As we will see, order-by-order contributions of the employed method are not manifestly gauge invariant. 
But they combine in the end and provide gauge invariant results.
In this paper we will discuss the case $\Omega\ne 1$ in $D=4$ in detail, for the interesting special case $\Omega=1$,
we refer to a subsequent paper \cite{Grosse:2007xx} and to a recent conference report \cite{Grosse:2006hh}.

In the following two sections, we describe our model and the employed method of extracting the singular contributions
of the one-loop action in detail. In Section~\ref{calc}, we sketch the explicit calculations. The results are summarised
in Subsection~\ref{res} and discussed in the final Section.


\section{\label{model}The Model}

Let us start from the action~(\ref{action})
\beb
S & = & \int d^4 x \left( 
\frac{1}{2} \phi \star [\tilde X_\nu,\, [\tilde X^\nu,\, \phi]_\star ]_\star 
+ \frac{\Omega^2}{2} 
\phi \star \{\tilde X^\nu , \{ \tilde X_\nu ,\phi\}_\star \}_\star 
\right.
\nonumber
\\
&&
+ \left. \frac{\mu^2}{2} \phi \star \phi 
+ \frac{\lambda}{4!} \phi\star \phi  \star \phi \star \phi\right)(x)\;.
\eeb
The expansion of $S$ yields
\bea
S & = & S_0 + \int d^4 x\,\frac{1}{2} \phi \star 
 \Big( 
2 \mathrm{i}  A^\nu \star \partial_\nu \phi 
- 2\mathrm{i} \partial_\nu \phi \star A^\nu 
\nonumber
\\
&& + 2 (1+\Omega^2) A_\nu \star A^\nu \star \phi 
- 2 (1-\Omega^2) A_\nu \star \phi \star A^\nu 
\nonumber
\\
&& + 2 \Omega^2 \{ \tilde x_\nu , (A^\nu \star \phi 
+ \phi \star A^\nu) \}_\star  \Big)\;,
\eea
where $S_0$ denotes the free part ot the action independent of $A$.
Now we compute the second derivative:
\bea
\frac{\delta^2 S}{ \delta \phi^2}(\psi) & = &
 \frac{2}{\theta} H^0 \psi +
\frac{\lambda}{3!} \big(\phi \star \phi \star \psi + 
\psi \star \phi \star \phi + 
\phi \star \psi \star \phi \big)
\nonumber
\\*
&& + \mathrm{i} \partial_\nu A^\nu \star \psi
- \mathrm{i} \psi \star \partial_\nu A^\nu
+2 \mathrm{i}  A^\nu \star \partial_\nu \psi 
- 2\mathrm{i} \partial_\nu \psi \star A^\nu 
\nonumber
\\
&& + (1+\Omega^2) A_\nu \star A^\nu \star \psi 
- 2 (1-\Omega^2) A_\nu \star \psi \star A^\nu
+ (1+\Omega^2) \psi \star A_\nu \star A^\nu 
\label{Spsi}
\\
\nonumber
&& + 2 \Omega^2 
\Bigg(
	\tilde x_\nu \cdot (A^\nu \star \psi + \psi \star A^\nu) 
	+ (\tilde x_\nu \cdot \psi) \star A^\nu + A^\nu \star (\tilde x_\nu \cdot \psi)
\Bigg),
\eea
where 
\be
H^0 = \frac{\theta}{2} 
\left( - \frac{\partial^2}{\partial x_\nu \partial x^\nu} + 
4 \Omega^2 \tilde x_\nu \tilde x^\nu + \mu^2 \right)\;.
\label{Schr}
\ee
The oscillator term is considered as a modification of the free theory. 
We use the the following parametrisation of $\theta_{\mu\nu}$:
\begin{equation*}
(\theta_{\mu\nu}) = \left(
\begin{array}{cccc}
0 & \theta &  & \\
-\theta & 0 & & \\
&&0&\theta      \\
&&-\theta&0
\end{array}\right), \quad
(\theta^{-1}_{\mu\nu}) = \left(
\begin{array}{cccc}
0 & -1/\theta &&\\
1/\theta & 0&&  \\
&&0&-1/\theta   \\
&&1/\theta&0    
\end{array}
\right). 
\end{equation*}
We expand the fields in the matrix base of the Moyal plane, 
\be
A^\nu(x) =\sum_{p,q \in \mathbb{N}^{2}} A^\nu_{pq} f_{pq}(x)\;, 
\phi(x) = \sum_{p,q \in \mathbb{N}^{2}} \phi_{pq} f_{pq}(x)\;, 
\psi(x) = \sum_{p,q \in \mathbb{N}^{2}} \psi_{pq} f_{pq}(x)\;. 
\ee
This choice of basis simplifies the calculations. In the end, we will again represent the results in the $x$-basis.
Useful properties of this basis (which we also use in the Appendix) are reviewed in the 
Appendix of \cite{Grosse:2003nw}.
Using Eqns~(\ref{app1}) and (\ref{app2}) from the Appendix, we obtain for (\ref{Spsi})
\bea
\frac{\delta^2 S}{ \delta \phi^2}(f_{mn})(x) 
&=& \sum_{r,s \in \mathbb{N}^2} G_{rs;mn} f_{sr}(x)
\nonumber
\\
&& \hspace{-1.5cm}
+ \sum_{r \in \mathbb{N}^2} \Big(
\frac{\lambda}{3!} \phi \star \phi
+ (1+\Omega^2)\big( \tilde X_\nu\star \tilde X^\nu -\tilde x^2 \big) \Big)_{rm} f_{rn}(x) 
\nonumber
\\
&&\hspace{-1.5cm}
+ \sum_{s \in \mathbb{N}^2} \Big(
\frac{\lambda}{3!} \phi \star \phi
+ (1+\Omega^2) \big(\tilde X_\nu \star \tilde X^\nu - \tilde x^2 \big) \Big)_{ns} f_{ms}(x) 
\nonumber
\\
&& \hspace{-1.5cm}
+ \sum_{r,s \in \mathbb{N}^2} 
\Big(\frac{\lambda}{3!} \phi_{rm} \phi_{ns} 
- 2 (1-\Omega^2) A_{\nu,rm} A^\nu_{ns} \Big) f_{rs}(x) 
\nonumber
\\
&&\hspace{-1.5cm}
+ (1-\Omega^2)\mathrm{i}  \sqrt{\frac{2}{\theta}} 
\sum_{r \in \mathbb{N}^2} \Big(
\sqrt{n^1} A^{(1+)}_{\stackrel{r^1}{r^2}\stackrel{m^1}{m^2}}
f_{\stackrel{r^1}{r^2}\stackrel{n^1-1}{n^2}}
- \sqrt{n^1+1}A^{(1-)}_{\stackrel{r^1}{r^2}\stackrel{m^1}{m^2}}
f_{\stackrel{r^1}{r^2}\stackrel{n^1+1}{n^2}}
\nonumber
\\
&&  \hspace*{.5cm}  
+ \sqrt{n^2} A^{(2+)}_{\stackrel{r^1}{r^2}\stackrel{m^1}{m^2}}
f_{\stackrel{r^1}{r^2}\stackrel{n^1}{n^2-1}}  
- \sqrt{n^2+1} A^{(2-)}_{\stackrel{r^1}{r^2}\stackrel{m^1}{m^2}}
f_{\stackrel{r^1}{r^2}\stackrel{n^1}{n^2+1}}\Big)
\nonumber
\\
&& \hspace{-1.5cm}
- (1-\Omega^2)\mathrm{i} \sqrt{\frac{2}{\theta}} 
\sum_{s \in \mathbb{N}^2} \Big(
- \sqrt{m^1+1} A^{(1+)}_{\stackrel{n^1}{n^2}\stackrel{s^1}{s^2}} 
f_{\stackrel{m^1+1}{m^2}\stackrel{s^1}{s^2}}
+ \sqrt{m^1} A^{(1-)}_{\stackrel{n^1}{n^2}\stackrel{s^1}{s^2}} 
f_{\stackrel{m^1-1}{m^2}\stackrel{s^1}{s^2}}
\nonumber
\\
&& \hspace*{.5cm} 
- \sqrt{m^2+1} A^{(2+)}_{\stackrel{n^1}{n^2}\stackrel{s^1}{s^2}} 
f_{\stackrel{m^1}{m^2+1}\stackrel{s^1}{s^2}}
+ \sqrt{m^2} A^{(2-)}_{\stackrel{n^1}{n^2}\stackrel{s^1}{s^2}} 
f_{\stackrel{m^1}{m^2-1}\stackrel{s^1}{s^2}} \Big)\;,
\label{spp}
\eea
where 
\be
A^{(1\pm)}= A^1\pm \mathrm{i} A^2\;,\qquad
A^{(2\pm)}= A^3\pm \mathrm{i} A^4\;.
\ee
We extract the $(lk)$-component of (\ref{spp}):
\be
\frac{\theta}{2} 
\left(\frac{\delta^2 S}{ \delta \phi^2}(f_{mn})\right)_{lk} = 
H^0_{kl;mn} + \frac{\theta}{2} V_{kl;mn} \equiv H_{kl;mn}\;,
\label{SHB}
\ee
where 
\begin{align}
H^0_{mn;kl} 
&= \big(\frac{\mu^2\theta}{2} {+} (1{+}\Omega^2)
  (n^1{+}m^1{+}1) {+} (1{+}\Omega^2)(n^2{+}m^2{+}1) \big)
  \delta_{n^1k^1} \delta_{m^1l^1} \delta_{n^2k^2} \delta_{m^2l^2} \nonumber
  \\*
  & - (1{-}\Omega^2) \big(\sqrt{k^1l^1}\,
  \delta_{n^1+1,k^1}\delta_{m^1+1,l^1 } 
+ \sqrt{m^1n^1}\, \delta_{n^1-1,k^1}
  \delta_{m^1-1,l^1} \big) \delta_{n^2k^2} \delta_{m^2l^2} \nonumber
  \\*
  & - (1{-}\Omega^2) \big(\sqrt{k^2l^2}\,
  \delta_{n^2+1,k^2}\delta_{m^2+1,l^2 } 
+ \sqrt{m^2n^2}\, \delta_{n^2-1,k^2}
  \delta_{m^2-1,l^2} \big) \delta_{n^1k^1} \delta_{m^1l^1}
\label{G4D}
\end{align}
is the field-independent part and 
\begin{align}
V_{kl;mn} 
&= \Big(
\frac{\lambda}{3!} \phi \star \phi
+ (1+\Omega^2)\big(\tilde X_\nu \star \tilde X^\nu -\tilde x^2 \big) \Big)_{lm} \delta_{nk}
\nonumber
\\*
&+ \Big(
\frac{\lambda}{3!} \phi \star \phi
+ (1+\Omega^2) \big(\tilde X_\nu \star \tilde X^\nu -\tilde x^2 \big) \Big)_{nk} \delta_{ml}
\nonumber
\\
& + 
\Big(\frac{\lambda}{3!} \phi_{lm} \phi_{nk} 
- 2 (1-\Omega^2) A_{\nu,lm} A^\nu_{nk} \Big) 
\nonumber
\\
&+ (1-\Omega^2)\mathrm{i}  \sqrt{\frac{2}{\theta}} \Big(
\sqrt{n^1} A^{(1+)}_{\stackrel{l^1}{l^2}\stackrel{m^1}{m^2}}
\delta_{\stackrel{k^1}{k^2}\stackrel{n^1-1}{n^2}}
- \sqrt{n^1+1}A^{(1-)}_{\stackrel{l^1}{l^2}\stackrel{m^1}{m^2}}
\delta_{\stackrel{k^1}{k^2}\stackrel{n^1+1}{n^2}}
\nonumber
\\*
&  \hspace*{7em}  
+ \sqrt{n^2} A^{(2+)}_{\stackrel{l^1}{l^2}\stackrel{m^1}{m^2}}
\delta_{\stackrel{k^1}{k^2}\stackrel{n^1}{n^2-1}}  
- \sqrt{n^2+1} A^{(2-)}_{\stackrel{l^1}{l^2}\stackrel{m^1}{m^2}}
\delta_{\stackrel{k^1}{k^2}\stackrel{n^1}{n^2+1}}\Big)
\nonumber
\\
& 
- (1-\Omega^2)\mathrm{i} \sqrt{\frac{2}{\theta}} \Big(
- \sqrt{m^1+1} A^{(1+)}_{\stackrel{n^1}{n^2}\stackrel{k^1}{k^2}} 
\delta_{\stackrel{m^1+1}{m^2}\stackrel{l^1}{l^2}}
+ \sqrt{m^1} A^{(1-)}_{\stackrel{n^1}{n^2}\stackrel{k^1}{k^2}} 
\delta_{\stackrel{m^1-1}{m^2}\stackrel{l^1}{l^2}}
\nonumber
\\*
& \hspace*{7em} 
- \sqrt{m^2+1} A^{(2+)}_{\stackrel{n^1}{n^2}\stackrel{k^1}{k^2}} 
\delta_{\stackrel{m^1}{m^2+1}\stackrel{l^1}{l^2}}
+ \sqrt{m^2} A^{(2-)}_{\stackrel{n^1}{n^2}\stackrel{k^1}{k^2}} 
\delta_{\stackrel{m^1}{m^2-1}\stackrel{l^1}{l^2}} \Big)\;.
\label{sppc}
\end{align}
%
%
The heat kernel $e^{-tH^0}$ of the Schr\"odinger operator (\ref{Schr})
can be calculated from the propagator given in \cite{Grosse:2004yu}. In the matrix base
of the Moyal plane, it has the following representation:
\bea
\left( e^{-tH^0}\right)_{mn;kl} & = & e^{-2t\sigma^2}
\delta_{m+k,n+l} \prod_{i=1}^{2} K_{m^in^i;k^il^i}(t)\;,
\\
K_{m,m+\alpha;l+\alpha,l}(t) & = & \sum_{u=0}^{\textrm {min}(m,l)} 
\sqrt{\binom{m}{u}\binom{l}{u}
  \binom{\alpha+m}{m-u}\binom{\alpha+l}{l-u}}
\nonumber
\\
&& \times 
\frac{e^{-4\Omega t(\frac{1}{2} \alpha + u)}
(1-e^{-4\Omega t})^{m+l-2u}}{
(1-\frac{(1-\Omega)^2}{(1+\Omega)^2} e^{-4\Omega t})^{\alpha +m+l+1}}
\Big(\frac{4\Omega}{(1+\Omega)^2}\Big)^{\alpha+2u+1} 
\Big(\frac{1-\Omega}{1+\Omega}\Big)^{m+l-2u} 
\label{comp}\\
& = & \sum_{u=0}^{\textrm {min}(m,l)} 
\sqrt{\binom{m}{u}\binom{l}{u}
  \binom{\alpha+m}{m-u}\binom{\alpha+l}{l-u}} \\
\nonumber
& &\times \, 
e^{2 \Omega t} \left( \frac{1-\Omega^2}{2\Omega}\sinh (2\Omega t) \right)^{m+l-2u}
X_\Omega(t)^{\alpha+m+l+1}
\;,
\eea
where $2 \sigma^2 =(\mu^2\theta/2+ 4 \Omega )$, and we have defined
\begin{align}
\label{def1}
X_\Omega(t)= \frac{4\Omega}{
(1+\Omega)^2e^{2\Omega t}-(1-\Omega)^2 e^{-2\Omega t}} \; .
\end{align}

For $\Omega=1$, 
the interaction part of the action  simplifies a lot,
\bea
V_{kl;mn} 
& = & \Big(
\frac{\lambda}{3!} \phi \star \phi
+ 2\big( \tilde X_\mu \star \tilde X^\mu - \tilde x^2
\big) \Big)_{lm} \delta_{nk}
\nonumber
\\*
&& + \Big(
\frac{\lambda}{3!} \phi \star \phi
+ 2 \big( \tilde X_\mu \star \tilde X^\mu - \tilde x^2
\big) \Big)_{nk} \delta_{ml}
+ \frac{\lambda}{3!} \phi_{lm} \phi_{nk}
\label{om1}\\
& \equiv & a_{lm} \delta_{nk} + a_{nk}\delta_{ml} + \frac{\lambda }{3!} \phi_{lm}
\phi_{nk}\, ,
\eea
and for the heat kernel, we obtain the following simple expression:
\bea
\left( e^{-tH^0} \right)_{mn;kl} & = & \delta_{ml} \delta_{kn} e^{-2t\sigma^2} 
\prod_{i=1}^{2} e^{-2t(m^i + n^i)},\\
K_{mn;kl}(t) & = & \delta_{ml} \prod_{i=1}^{2}e^{-2t(m^i+k^i)},
\eea
where $\sigma^2=\frac{\mu^2\theta}4+2$.


\section{Method}

Given an operator $P$ on the algebra, we write
\be
P f_{mn} = \sum_{k,l} (P_{mn})_{lk} f_{lk} = \sum_{k,l} 
f_{lk} P_{kl;mn}\;.
\ee
Then, the composition of two such operators $P,Q$ reads 
\bea
P Q f_{mn} &= \sum_{k,l} (Q_{mn})_{lk} (P f_{lk}) 
= \sum_{k,l,r,s} (Q_{mn})_{lk} (P_{lk})_{rs} f_{rs} 
\nonumber
\\*
& = \sum_{k,l} (P f_{lk}) Q_{kl;mn}
= \sum_{k,l,r,s} f_{rs} P_{sr;lk} Q_{kl;mn}\;,
\eea
hence
\be
 [P Q]_{sr;mn}=\sum_{k,l} P_{sr;lk} Q_{kl;mn}\;.
\ee
The trace of such an operator is then given by 
\be
\mathrm{Tr} P = \sum_{m,n} P_{mn;nm}\;.  
\label{TrP}
\ee

Bearing in mind these index rules, we can compute the regularised one-loop effective action for the model defined
by the classical action (\ref{action}), which is given by 
\be
\Gamma^\epsilon_{1l}[\phi] = -\frac{1}{2} \int_\epsilon^\infty 
\frac{dt}{t} \,\mathrm{Tr}\left( e^{-t H} - e^{-t H^0} \right) \;.
\label{Gamma-e}
\ee
One way to proceed would be to use the Baker-Campbell-Hausdorff
formula (it is used e.g., in \cite{Gayral:2004cs}),
\be
\mathrm{Tr}\left(e^{-t H} -e^{-t H^0}\right) =
\mathrm{Tr} \left( \Big( -\frac{\theta}{2} t V + \frac{t^2}{2}
  \frac{\theta}{2} [ H^0,V] - \frac{t^3}{6} \frac{\theta}{2}
[H^0,[H^0,V]] + \frac{t^2}{2} \frac{\theta^2}{4} V^2 \Big) 
e^{-t H^0} \right)\;.
\label{BCH}
\ee
However, for reasons of convergence, we use the Duhamel formula instead. We have to iterate
the identity
\bea
e^{-tH}-e^{-tH^0} &=& \int_0^t d\sigma \; \frac{d}{d \sigma} \left(
e^{-\sigma H} e^{-(t-\sigma)H^0} \right)
\nonumber
\\
&=& -\int_0^t d\sigma \; 
e^{-\sigma H} \,\frac{\theta}{2} V \,e^{-(t-\sigma)H^0} \;,
\eea
giving 
\bea
e^{-tH} &=& e^{-t H^0} - \frac{\theta}{2} \int_0^t d t_1 
e^{-t_1 H^0} V e^{-(t-t_1) H^0} 
\nonumber
\\
&&+  \Big(\frac{\theta}{2}\Big)^2 
\int_0^t d t_1 \int_0^{t_1} d t_2 
e^{-t_2 H^0} V e^{-(t_1-t_2) H^0} V e^{-(t-t_1) H^0} + \dots
\eea
We thus obtain\footnote{In the $V$-bilinear integral we set
  $t_1-t_2=t-t'$ so that the $t_2$ integration goes from $0$ to $t'$.}
\bea
\label{duhamel-action}
\Gamma_{1l}^\epsilon & = & \frac{\theta}4 \int_\epsilon^\infty dt \textrm{ Tr }
    V e^{-tH^0} - \frac{\theta^2}8 \int_\epsilon^\infty \frac{dt}{t} \int_0^t dt'\, t'
    \textrm{ Tr } V e^{-t'H^0} V e^{-(t-t')H^0}\\
\nonumber
&& \hspace{-.8cm} +  \frac{\theta^3}{16} \int_\epsilon^\infty \frac{dt}{t} \int_0^t dt' \int_0^{t'} dt'' \, t''
    \textrm{ Tr } V e^{-t''H^0} V e^{-(t'-t'')H^0} V e^{-(t-t')H^0}\\
\nonumber
& & \hspace{-.8cm} - \frac{\theta^4}{32} \int_\epsilon^\infty \frac{dt}{t} \int_0^t dt' \int_0^{t'} dt'' 
    \int_0^{t''} dt''' \, t'''
    \textrm{ Tr } V e^{-t'''H^0} V e^{-(t''-t''')H^0} V e^{-(t'-t'')H^0} V e^{-(t''-t''')H^0}\\
\nonumber
& + & \mathcal O(\theta^5)
\\
\nonumber
& = & \Gamma^\epsilon_{1l,1}[\phi] + \Gamma^\epsilon_{1l,2}[\phi] + \Gamma^\epsilon_{1l,3}[\phi] + \Gamma^\epsilon_{1l,4}[\phi]+
\mathcal O(\theta^5)\,.
\eea
The first term in both expansions coincides. Contributions to Eqn.~(\ref{duhamel-action}) higher than fourth order are finite.

For simplicity, we introduce an additional double index notation:
\be
\prod_{i=1}^{2} K_{m^in^i;k^il^i}(t) \equiv K_{mn;kl}(t)
\ee
Indices not indexed by $1$ or $2$ are supposed to be double indices, unless otherwise stated.

Operators $H^0$ and $V$ entering the heat kernel obey obvious scaling relations. Defining
\bea
\nonumber
v & = & \frac{V}{1+\Omega^2},\\
\nonumber
h^0 & = & \frac{H^0}{1+\Omega^2},
\eea
and the auxiliary parameter $\tau$
$$
\tau = t\, (1+\Omega^2)\,,
$$
it leads to operators depending beside on $\theta$ only on the following three parameters:
\bea
\rho = \frac{1-\Omega^2}{1+\Omega^2},\\
\tilde \epsilon = \epsilon\, (1+\Omega^2),\\
\tilde \mu^2 = \frac{\mu^2\theta}{1+\Omega^2}.
\eea

The task of this paper is to extract the divergent contributions of the expansion~(\ref{duhamel-action}).
In order to do so, we expand the integrands for small auxiliary parameters. The divergencies are due to 
infinite sums over indices occurring in the heat kernel but not in the gauge field $A$.
After integrating over the auxiliary parameters, we obtain the divergent contributions listed and calculated 
in the next section. In the end, we convert the results to $x$-space using 
$$
\sum_m B_{mm} = \frac1{4\pi^2\theta^2} \int d^4x \,B(x),
$$
where $B(x) = \sum_{m,n} B_{mn} f_{mn}(x)$.


\section{\label{calc}Calculations}

We concentrate on the gauge fields and set $\lambda=0$. The $A^2$ term in Eq.~({\ref{sppc}) does not need to be
considered, since it leads to finite contributions in all orders. Let us examine the calculation of the Duhamel expansion 
(\ref{duhamel-action}) order by order.


\subsection{First order}

To first order the Duhamel expansion (\ref{duhamel-action}) of the effective action yields
\bea
\nonumber
\Gamma_{1l,1}^\epsilon & = & 
	\frac{\theta}4 \int_\epsilon^\infty dt  \sum_{k,l,m,n} V_{kl;mn} (e^{-tH^0})_{nm;lk}\\
\nonumber
& = & \frac{\theta}4 \int_\epsilon^\infty dt e^{-2t\sigma^2} \sum_{k,l,m,n} K_{nm;lk}(t)
\Bigg\{ 
\bigg(
\sqrt{n^1}A^{(1+)}_{lm} \delta_{\stackrel{n^1}{n^2}\stackrel{k^1+1}{k^2}}
-\sqrt{n^1+1}A^{(1-)}_{lm} \delta_{\stackrel{n^1+1}{n^2}\stackrel{k^1}{k^2}}\\
\nonumber
&&\hspace{1cm}
+\sqrt{n^2}A^{(2+)}_{lm} \delta_{\stackrel{n^1}{n^2}\stackrel{k^1}{k^2+1}}
-\sqrt{n^2+1}A^{(2-)}_{lm} \delta_{\stackrel{n^1}{n^2+1}\stackrel{k^1}{k^2}}
\\
\nonumber
&& \hspace{1cm}
+\sqrt{m^1+1}A^{(1+)}_{nk} \delta_{\stackrel{l^1}{l^2}\stackrel{m^1+1}{m^2}}
-\sqrt{m^1}A^{(1-)}_{nk} \delta_{\stackrel{l^1+1}{l^2}\stackrel{m^1}{m^2}}
\\
\nonumber
&& \hspace{1cm}
+\sqrt{m^2+1}A^{(2+)}_{nk} \delta_{\stackrel{l^1}{l^2}\stackrel{m^1}{m^2+1}}
-\sqrt{m^2}A^{(2-)}_{nk} \delta_{\stackrel{l^1}{l^2+1}\stackrel{m^1}{m^2}}
\bigg) i\sqrt{\frac2{\theta}}\, (1-\Omega^2)\\
\nonumber
&& 
+ (1+\Omega^2)
\bigg( (\tilde X_\nu \star \tilde X^\nu -\tilde x^2)_{lm}\delta_{nk} + (\tilde X_\nu \star \tilde X^\nu -\tilde x^2)_{nk}\delta_{lm} 
\bigg)
\Bigg\}\,.
\eea
The divergences are due to partial traces of the Kernel $K(t)$, i.e., sums over indices that occur in $K(t)$, but not
in the gauge fields. There are two relevant traces - no double index notation is implied here - , namely 
\bea
\nonumber
&&
\sum_{n=0}^\infty K_{mn;nm}(t) =
\\
\nonumber
& = &  \sum_{n=0}^\infty \sum_{v=0}^{\min(m,n)} 
\binom{m}{v} \binom{n}{v}
\frac{e^{-4\Omega t(\frac{1}{2}n + \frac{1}{2} m - v)}
(1-e^{-4\Omega t})^{2v}}{
(1-\frac{(1-\Omega)^2}{(1+\Omega)^2} e^{-4\Omega t})^{n+m+1}}
\Big(\frac{4\Omega}{(1+\Omega)^2}\Big)^{n+m-2v+1} 
\Big(\frac{1-\Omega}{1+\Omega}\Big)^{2v}\\
\nonumber
& = & e^{2\Omega t} X_\Omega(t)^{m+1}\Big( \sum_{n=0}^\infty X_\Omega(t)^n + m (1-\Omega^2)^2
\frac{(e^{2\Omega t} - e^{-2\Omega t})^2}{16\Omega^2} \sum_{n=0}^\infty n\, X_\Omega(t)^n \Big) 
+ \mathcal O(t) \\
\nonumber
& = &  e^{2\Omega t} \frac{ X_\Omega(t)^{m+1} } {1-X_\Omega(t) } \left( 
1+ m \frac{(1-\Omega^2)^2 (e^{2\Omega t} - e^{-2\Omega t})^2 X_\Omega(t)}
    {16 \Omega^2 (1-X_\Omega (t) ) } \right) + \mathcal O(t)
    \\
\label{trace1}
& = & \frac1{(1+\Omega^2)t} \left(
    1+ 2\Omega t (1-\frac{(2m+1)\Omega}{1+\Omega^2}) \right) +
    \mathcal O(t)
\eea
and
\bea
\label{trace2}
\sum_{n=0}^\infty \sqrt{n+1} K_{m+1,n+1;n,m}(t) =
\frac{\sqrt{m+1}}{t} \frac{1-\Omega^2}{(1+\Omega^2)^2} 
\left(1+2 \Omega t 
\, ( 1- \frac{ 2(m+1) \Omega}{1+\Omega^2})
 \right) + \mathcal{O}(t) \;.
\eea
The partial traces together with Eqns.~(\ref{app2}) and (\ref{matrix2-4}) and the identity
$$
\{ \tilde x^\mu, A_\mu\}_\star = 2 \tilde x^\mu A_\mu = 2(\tilde x A)
$$
lead to the following result:
\bea
\Gamma^\epsilon_{1l,1} & = &
\frac{-1}{12\pi^2} \int d^4x \Bigg\{ 
\frac{-1}{\tilde \epsilon} 
	\left(
	\frac{3}{2\theta} A_\nu \star A^\nu
        + \frac{3(1-\rho^2)}{2 \theta} (2 \tilde x A) \right) 
\\
\nonumber
&& - \frac{\tilde \mu^2}{\theta} \ln\epsilon \left(
        \frac34 A_\nu \star A^\nu
        + \frac{3(1-\rho^2)}{4} (2 \tilde x A) \right)
\\
\nonumber
&& - \ln\epsilon \frac{3(1-\rho^2)}4 \tilde x^2 (A_\mu\star A^\mu) + \ln\epsilon 
		\frac{3\rho^2(1-\rho^2)}{4\theta} (2 \tilde x A) - 
		\ln\epsilon \frac{3(1-\rho^2)^2}{2}  \tilde x^2 (\tilde x A)
\Bigg\}
\label{1st-order}\\
\nonumber
&& + \mathcal O(\epsilon^0)\;.
\eea
Both, logarithmic and quadratic divergences occur. Some logarithmic divergences also stem from the 
constant term in the expansion of the partial traces (\ref{trace1}) and (\ref{trace2}), which will be called 
{\it subleading divergences}.


\subsection{Second order}

The second order calculations are quite involved, and there are numerous contributions.
Thus, for the clarity of presentation, we divide calculations up. According to Eq.~(\ref{duhamel-action}), we
need to calculate
\be
\label{2nd}
\Gamma_{1l,2}^\epsilon = -\frac{\theta^2}8 
\int_\epsilon^\infty \frac{dt}{t} e^{-2\sigma^2 t}
\int_0^t dt'\, t' \sum_{k,l,m,n,a,b,c,d} V_{kl;mn} K(t')_{nm;ab} V_{ba;cd} K(t-t')_{dc;lk}\,,
\ee
where the potential $V$ is given by Eq.~(\ref{sppc}). Let us rewrite Eq.~(\ref{sppc}) in a schematic way:
\be
V \,\, "=" \,\, \tilde X^2 + A,
\ee
where the part "$A$" consists of two different blocks, later on referred to as first and second block.
To second order, we have to consider two potentials. Therefore, there are three different contributions which
all produce divergent terms.

\subsubsection{$\tilde X^2-\tilde X^2$}

First, we insert for both potentials the terms proportional to $\tilde X\star \tilde X - \tilde x^2$.
We obtain

\bea
\nonumber
\Gamma_{1l,2.1}^\epsilon & = & - \frac{\theta^2}8 
    \int_\epsilon^\infty \frac{dt}{t} e^{-2\sigma^2 t}\int_0^t dt'\, t'
    K(t')_{nm;mn} K(t-t')_{kl;lk} (1+\Omega^2)^2 
    \\
\nonumber
&& \times ((\tilde X^{\star 2}-\tilde x^2)_{lm} \delta_{nk}
    + (\tilde X^{\star 2}-\tilde x^2)_{nk} \delta_{lm})
    ((\tilde X^{\star 2}-\tilde x^2)_{ml} \delta_{kn}
    + (\tilde X^{\star 2}-\tilde x^2)_{kn} \delta_{ml})\\
\nonumber
&& + \mathcal O(\epsilon^0)\\
\nonumber
& = & - 2 (1+\Omega^2)^2 \frac{\theta^2}8 
    \int_\epsilon^\infty \frac{dt}{t} e^{-2\sigma^2 t}\int_0^t dt'\, t'
    K(t')_{nm;mn} K(t-t')_{kl;lk}
\\
\nonumber
&& \hspace{1cm} \times (\tilde X^{\star 2}-\tilde x^2)_{lm} 
    (\tilde X^{\star 2}-\tilde x^2)_{ml} + \mathcal O(\epsilon^0)\\
\nonumber
& = & - (1+\Omega^2)^2 \frac{\theta^2}4 \frac1{4\pi^2\theta^2} \int d^4x\,  
    \int_\epsilon^\infty \frac{dt}{t} e^{-2\sigma^2 t}\int_0^t dt'\, t'
    \frac1{t^2(1+\Omega^2)^2} (\tilde X^{\star 2}-\tilde x^2)^{\star 2}
 + \mathcal O(\epsilon^0)\\
\nonumber
& = &  \frac{\ln\epsilon}{4\pi^2} \int d^4x\,  
    \frac18 (\tilde X^{\star 2}-\tilde x^2)^{\star 2} + \mathcal O(\epsilon^0)\\
\nonumber
& = &  \frac{\ln\epsilon}{12\pi^2(1+\Omega^2)^2} \int d^4x\,  
    \frac38 (1+\Omega^2)^2 
    \Big\{ (\tilde X^{\star 2})^{\star2}-(\tilde x^2)^2 - 2\tilde x^2 (A_\mu \star
    A^\mu)\\
&& - 4 \tilde x^2 (\tilde x A) \Big\} + \mathcal O(\epsilon^0)\,,   
\eea
where we have used Eq.~(\ref{trace1}). Summation over all indices is implied.

\subsubsection{$A-\tilde X^2$}

Let us examine the field content $A^{(-1)}-\tilde X^2$, where $A$ is taken from the first block. 
For this contribution to the effective action 
is given by the following expression:
\bea
\nonumber
\Gamma_{1l,2.2}^\epsilon & = & \frac{i \theta^2}8 \sqrt{\frac2{\theta}} (1-\Omega^2)(1+\Omega^2)
\int_\epsilon^\infty \frac{dt}{t} e^{-2\sigma^2 t} \int_0^t dt' t'\, 
\sqrt{n^1+1} \delta_{\stackrel{n^1+1}{n^2} \stackrel{k^1}{k^2} } A^{(1-)}_{lm}\\
\nonumber
&&\times
\bigg(
(B^2-\tilde x^2)_{ac} \delta_{db} + (\tilde X^{\star2}-\tilde x^2)_{db} \delta_{ac}
\bigg)
K_{nm;ab}(t') K_{dc;lk}(t-t') \delta_{n+a,m+b} \delta_{d+l,c+k}\\
\nonumber
& = & \frac{i \theta^2}8 \sqrt{\frac2{\theta}} (1-\Omega^2)(1+\Omega^2)
\int_\epsilon^\infty \frac{dt}{t} e^{-2\sigma^2 t} \int_0^t dt' t'\,
\\
\label{e113}
&& \hspace{-1cm}
\times \Bigg(
\sqrt{n^1+1} A^{(1-)}_{lm} (\tilde X^{\star2}-\tilde x^2)_{ac} K_{nm;ad}(t') 
	K_{\stackrel{d^1}{d^2}\stackrel{c^1}{c^2};\stackrel{l^1}{l^2}\stackrel{n^1+1}{n^2}}(t-t') 
	\delta_{n+a,m+d} \delta_{\stackrel{d^1+l^1}{d^2+l^2},\stackrel{c^1+n^1+1}{c^2+n^2}}
\\
\label{e114}
&& \hspace{-.8cm}
+ \sqrt{n^1+1} A^{(1-)}_{lm} (\tilde X^{\star 2}-\tilde x^2)_{db} K_{nm;ab}(t') 
	K_{\stackrel{d^1}{d^2}\stackrel{a^1}{a^2};\stackrel{l^1}{l^2}\stackrel{n^1+1}{n^2}}(t-t') 
	\delta_{n+a,m+b} \delta_{\stackrel{d^1+l^1}{d^2+l^2},\stackrel{a^1+n^1+1}{a^2+n^2}}
\Bigg)\,.
\eea
The contribution of line~(\ref{e114}) is finite, and line~(\ref{e113}) gives two divergent contributions:
\bea
\Gamma_{1l,2.2}^\epsilon & = & \frac{i \theta^2}8 \sqrt{\frac2{\theta}} (1-\Omega^2)(1+\Omega^2)
\int_\epsilon^\infty \frac{dt}{t} e^{-2\sigma^2 t} \int_0^t dt' t'\, 
\\
\nonumber
&& 
\times \Bigg(
\sqrt{n^1+1} A^{(1-)}_{cm} (\tilde X^{\star 2}-\tilde x^2)_{\stackrel{m^1+1}{m^2}\stackrel{c^1}{c^2}} 
	K_{\stackrel{n^1}{n^2}\stackrel{m^1}{m^2};\stackrel{m^1+1}{m^2}\stackrel{n^1+1}{n^2}}(t') 
	K_{\stackrel{n^1+1}{n^2}\stackrel{c^1}{c^2};\stackrel{c^1}{c^2}\stackrel{n^1+1}{n^2}}(t-t') 
\\
\nonumber	
&& \hspace{1cm}
+ \sqrt{n^1+1} A^{(1-)}_{\stackrel{c^1+1}{c^2}\stackrel{m^1}{m^2}} (\tilde X^{\star 2}-\tilde x^2)_{mc} 
	K_{nm;mn}(t') 
	K_{\stackrel{n^1}{n^2}\stackrel{c^1}{c^2};\stackrel{c^1+1}{c^2}\stackrel{n^1+1}{n^2}}(t-t') 
\Bigg)
+\mathcal O(\epsilon^0)\,.
\eea
The formulae for the partial traces over two kernels are given in the Appendix, Eqns.~(\ref{app4})
and (\ref{app5}), resp.
Only the leading terms in the expansions are necessary, since the subleading terms are already finite.
\bea
\nonumber
\Gamma_{1l,2.2}^\epsilon & = & \frac{i \theta^2}8 \sqrt{\frac2{\theta}} (1-\Omega^2)(1+\Omega^2)
\int_\epsilon^\infty \frac{dt}{t} e^{-2\sigma^2 t} \int_0^t dt' t'\, 
\\
\nonumber
&& 
\times \Bigg(
\sqrt{m^1+1} A^{(1-)}_{cm} (B^2-\tilde x^2)_{\stackrel{m^1+1}{m^2}\stackrel{c^1}{c^2}} 
	\frac1{t(1+\Omega^2)} \frac{1-\Omega^2}{t^2(1+\Omega^2)^2} t' 
\\
\nonumber	
&& \hspace{1cm}
+ \sqrt{m^1+1} A^{(1-)}_{\stackrel{c^1+1}{c^2}\stackrel{m^1}{m^2}} (B^2-\tilde x^2)_{mc} 
	 \frac1{t(1+\Omega^2)} \frac{1-\Omega^2}{t^2(1+\Omega^2)^2} (t-t')
\Bigg)
+\mathcal O(\epsilon^0)\\
& = & -i\frac{\theta^2}{24} \sqrt{\frac2{\theta}} \rho^2 \frac{\ln\epsilon}4 \sqrt{m^1+1}
\\
\nonumber
&& \times \Bigg(
2 A^{(1-)}_{cm} A^{(1+)}_{\stackrel{m^1+1}{m^2}\stackrel{a^1}{a^2}} A^{(1-)}_{ac} +
2 A^{(1-)}_{cm} A^{(1-)}_{\stackrel{m^1+1}{m^2}\stackrel{a^1}{a^2}} A^{(1+)}_{ac} +
  A^{(1-)}_{\stackrel{m^1+1}{m^2}\stackrel{c^1}{c^2}} A^{(1-)}_{ca} A^{(1+)}_{am} 
\\
\nonumber
&& \hspace{1cm} 
+
A^{(1-)}_{\stackrel{m^1+1}{m^2}\stackrel{c^1}{c^2}} A^{(1+)}_{ca} A^{(1-)}_{am} +
4 A^{(1-)}_{cm} (2\tilde x A)_{\stackrel{m^1+1}{m^2}\stackrel{c^1}{c^2}} +
2 A^{(1-)}_{\stackrel{m^1+1}{m^2}\stackrel{c^1}{c^2}} (2\tilde x A)_{cm}
\Bigg)
+\mathcal O(\epsilon^0)\,.
\eea  
We also need to consider the configuration $\tilde X^2-A^{(-1)}$, where $A^{(-1)}$ is on the second position.
The result is similar to the one above, we obtain
\bea
\nonumber
\Gamma_{1l,2.3}^\epsilon & = & -i\frac{\theta^2}{24} \sqrt{\frac2{\theta}} \rho^2 \frac{\ln\epsilon}4 \sqrt{m^1+1}
\\
\nonumber
&& \times \Bigg(
  A^{(1-)}_{cm} A^{(1+)}_{\stackrel{m^1+1}{m^2}\stackrel{a^1}{a^2}} A^{(1-)}_{ac} +
  A^{(1-)}_{cm} A^{(1-)}_{\stackrel{m^1+1}{m^2}\stackrel{a^1}{a^2}} A^{(1+)}_{ac} +
2 A^{(1-)}_{\stackrel{m^1+1}{m^2}\stackrel{c^1}{c^2}} A^{(1-)}_{ca} A^{(1+)}_{am} 
\\
\nonumber
&& \hspace{1cm} 
+
2 A^{(1-)}_{\stackrel{m^1+1}{m^2}\stackrel{c^1}{c^2}} A^{(1+)}_{ca} A^{(1-)}_{am} +
2 A^{(1-)}_{cm} (2\tilde x A)_{\stackrel{m^1+1}{m^2}\stackrel{c^1}{c^2}} +
4 A^{(1-)}_{\stackrel{m^1+1}{m^2}\stackrel{c^1}{c^2}} (2\tilde x A)_{cm}
\Bigg)
+\mathcal O(\epsilon^0)
\eea
From the second block, we obtain the same results as above. 
Using Eqns.~(\ref{matrix2-1}) and (\ref{matrix2}) and taking into account the contributions from the second
oscillator, we obtain
\bea
\label{calcABB}
\Gamma_{1l,2.4}^\epsilon & = & \frac{-\ln\epsilon}{12 \pi^2 (1+\Omega^2)^2} \int _\epsilon ^\infty d^4x\, 
(1-\Omega^2)^2 \Bigg(
\frac34  A_\mu \star A^\mu\star \{\tilde x_\nu, \star A^\nu\}_\star 
\\
\nonumber
&& \hspace{2cm}
+ \frac34 \{\tilde x_\nu , A^\nu\}_\star \star \{\tilde x_\mu , A^\mu \}_\star \Bigg)\,.
\eea

\subsubsection{$A-A$}

Divergent contributions are built from fields $A^{(1+)}$ and 
$A^{(1-)}$, resp., $A^{(2+)}$ and $A^{(2-)}$ from the same block. 
Plus and minus need not to be saturated. Mixed contributions 
containing fields from both oscillators are finite.


\paragraph{$\mathbf{A^{(1+)}-A^{(1+)}}$.}

Let us consider contributions with the field content $A^{(1+)}$--$A^{(1+)}$, from the first block.
From Eq.~(\ref{2nd}) we obtain
\bea
\nonumber
\nonumber
\Gamma_{1l,2.5}^\epsilon & = & \frac{\theta}4 \int_\epsilon^\infty \frac{dt}{t} e^{-2t\sigma^2} \int_0^t
dt' t' \,   (1-\Omega^2)^2 \sqrt{n^1} \sqrt{d^1} \, 
A^{(1+)}_{lm} A^{(1+)}_{ac} K_{nm;ab}(t') K_{dc;lk}(t-t')\\
&& \hspace{1cm}
\times \delta_{\stackrel{k^1+1}{k^2}\stackrel{n^1}{n^2}}\, \delta_{\stackrel{b^1+1}{b^2}\stackrel{d^1}{d^2}}\, 
    \delta_{n+a,m+b} \, \delta_{d+l,c+k}
\\
\nonumber
& = &
 \frac{\theta}4 \int_\epsilon^\infty \frac{dt}{t} e^{-2t\sigma^2} \int_0^t
dt' t' \,   (1-\Omega^2)^2 \sqrt{k^1+1} \sqrt{d^1+1} \, A^{(1+)}_{lm} A^{(1+)}_{ac}\\
\nonumber
&& 
\times
K_{\stackrel{k^1+1}{k^2},\stackrel{m^1}{m^2};\stackrel{a^1}{a^2},\stackrel{b^1}{b^2}}(t') 
K_{\stackrel{b^1+1}{b^2},\stackrel{c^1}{c^2};\stackrel{l^1}{l^2},\stackrel{k^1}{k^2}}(t-t') 
\\
\nonumber
& = & \frac{\theta}4 \int_\epsilon^\infty \frac{dt}{t} e^{-2t\sigma^2} \int_0^t
dt' t' \,   (1-\Omega^2)^2 \\
\nonumber
&& \times \Bigg\{
(k^1+1) A^{(1+)}_{\stackrel{l^1}{l^2}\stackrel{m^1+1}{m^2}} A^{(1+)}_{\stackrel{m^1}{m^2}\stackrel{l^1+1}{l^2}} 
K_{\stackrel{k^1+1}{k^2},\stackrel{m^1+1}{m^2};\stackrel{m^1}{m^2},\stackrel{k^1}{k^2}}(t') 
K_{\stackrel{k^1+1}{k^2},\stackrel{l^1+1}{l^2};\stackrel{l^1}{l^2},\stackrel{k^1}{k^2}}(t-t')\\
\nonumber
&&+
\sqrt{k^1+1}\sqrt{k^1+2}\,  A^{(1+)}_{lm} A^{(1+)}_{\stackrel{m^1}{m^2}\stackrel{l^1+2}{l^2}}
K_{\stackrel{k^1+1}{k^2},\stackrel{m^1}{m^2};\stackrel{m^1}{m^2},\stackrel{k^1+1}{k^2}}(t') 
K_{\stackrel{k^1+2}{k^2},\stackrel{l^1+2}{l^2};\stackrel{l^1}{l^2},\stackrel{k^1}{k^2}}(t-t')\\
\nonumber
&&+
\sqrt{k^1+1}\sqrt{k^1+2} \, A^{(1+)}_{\stackrel{m^1}{m^2}\stackrel{l^1+2}{l^2}} A^{(1+)}_{lm} 
K_{\stackrel{k^1+2}{k^2},\stackrel{l^1+2}{l^2};\stackrel{l^1}{l^2},\stackrel{k^1}{k^2}}(t') 
K_{\stackrel{k^1+1}{k^2},\stackrel{m^1}{m^2};\stackrel{m^1}{m^2},\stackrel{k^1+1}{k^2}}(t-t')
\Bigg\}
\\
\nonumber
&& +\mathcal O(\epsilon^0)\,.
\eea
The other contractions of the two kernels yield finite contributions. In the other cases, 
the difference in indices is two or bigger, and we have for small $t$
\bea
\nonumber
K_{m+\beta ,n+\beta ;n,m}(t)  & = & \sum_v \left(
\binom{m+\beta}{v+\beta} \binom{m}{v} \binom{n+\beta}{v+\beta} \binom{n}{v} \right)^{1/2} e^{2\Omega t}
\\
\nonumber
&& \times 
\left( \frac{1-\Omega^2}{2\Omega} \sinh (2\Omega t) \right)^{2v+\beta} X_\Omega(t)^{m+n+\beta+1}\\
\nonumber
& = &
\left(
\binom{m+\beta}{\beta} \binom{n+\beta}{\beta} \right)^{1/2} (1-\Omega^2)^\beta t^\beta
X_\Omega (t)^{m+n}(1 +\mathcal O(t))\,,
\eea
with $\beta\in \mathbb N$ fixed. With increasing $\beta$ the results are less divergent, for $\beta=3$
already finite. We get
\bea
\nonumber
\Gamma_{1l,2.5}^\epsilon & = & - \frac{\theta}{24} \left( \frac{1-\Omega^2}{1+\Omega^2} \right)^4 \ln\epsilon \sqrt{m^1+1}\sqrt{c^1+1}\,
A^{(1+)}_{\stackrel{c^1}{c^2}\stackrel{m^1+1}{m^2}} A^{(1+)}_{\stackrel{m^1}{m^2}\stackrel{c^1+1}{c^2}}\\
\label{1++}
&&
- \frac{\theta}{12} \left( \frac{1-\Omega^2}{1+\Omega^2} \right)^4 \ln\epsilon \,\sqrt{c^1}\sqrt{c^1+1} \,
A^{(1+)}_{\stackrel{c^1-1}{c^2}\stackrel{m^1}{m^2}} A^{(1+)}_{\stackrel{m^1}{m^2}\stackrel{c^1+1}{c^2}}\\
\nonumber
&& +\mathcal O(\epsilon^0)\,.
\eea
The same contribution comes from the second block. Therefore, there is an overall factor of 2.


\paragraph{$\mathbf{A^{(1-)}-A^{(1-)}}$.}

Similarly, for $A^{(1-)}$--$A^{(1-)}$ - first block - we obtain:
\bea
\nonumber
\Gamma_{1l,2.6}^\epsilon & = & \frac{\theta}4 \int_\epsilon^\infty \frac{dt}{t} e^{-2t\sigma^2} \int_0^t
dt' t' \,   (1-\Omega^2)^2 \sqrt{n^1} \sqrt{d^1} \, 
A^{(1-)}_{lm} A^{(1-)}_{ac} K_{nm;ab}(t') K_{dc;lk}(t-t')\\
\nonumber
&& \hspace{1cm}
\times \delta_{\stackrel{k^1}{k^2}\stackrel{n^1+1}{n^2}}\, \delta_{\stackrel{b^1}{b^2}\stackrel{d^1+1}{d^2}}\, 
    \delta_{n+a,m+b} \, \delta_{d+l,c+k}
\\
\label{1--}
& = & - \frac{\theta}{24} \left( \frac{1-\Omega^2}{1+\Omega^2} \right)^4 \ln\epsilon \sqrt{m^1+1}\sqrt{c^1+1}\,
A^{(1-)}_{\stackrel{c^1+1}{c^2}\stackrel{m^1}{m^2}} A^{(1-)}_{\stackrel{m^1+1}{m^2}\stackrel{c^1}{c^2}}\\
\nonumber
&&
- \frac{\theta}{12} \left( \frac{1-\Omega^2}{1+\Omega^2} \right)^4 \ln\epsilon \,\sqrt{c^1}\sqrt{c^1+1} \,
A^{(1-)}_{\stackrel{c^1+1}{c^2}\stackrel{m^1}{m^2}} A^{(1-)}_{\stackrel{m^1}{m^2}\stackrel{c^1-1}{c^2}}
+\mathcal O(\epsilon^0)\,.
\eea
Again, we have to take into account an overall factor of 2, which
results from the equal contribution from the second block.

\paragraph{$\mathbf{A^{(1+)}-A^{(1-)}}$.}

Next, let us consider the contribution $A^{(1+)}-A^{(1-)}$ from the first block:
\bea
\nonumber
\Gamma_{1l,2.7}^\epsilon & = &
- \frac{\theta}4 \int_\epsilon^\infty \frac{dt}{t} \int_0^t
dt' t' \, e^{-2t\sigma^2}  (1-\Omega^2)^2 \sqrt{n^1}
\sqrt{d^1+1} \, A^{(1+)}_{lm} A^{(1-)}_{ac} K_{nm;ab}(t') K_{dc;lk}(t-t')\\
\nonumber
&& \hspace{1cm}
\times \delta_{\stackrel{k^1+1}{k^2}\stackrel{n^1}{n^2}}\, \delta_{\stackrel{b^1}{b^2}\stackrel{d^1+1}{d^2}}\, 
    \delta_{n+a,m+b} \, \delta_{d+l,c+k}
\\
\nonumber
& = & - \frac{\theta}4 \int_\epsilon^\infty \frac{dt}{t} \int_0^t
dt' t' \, e^{-2t\sigma^2}  (1-\Omega^2)^2\\
&& \label{contraction-I}
\times
\Bigg\{
(k^1+1)
A^{(1+)}_{cm} A^{(1-)}_{mc} 
K_{\stackrel{k^1+1}{k^2},\stackrel{m^1}{m^2};\stackrel{m^1}{m^2},\stackrel{k^1+1}{k^2}}(t') 
K_{\stackrel{k^1}{k^2}\stackrel{c^1}{c^2};\stackrel{c^1}{c^2}\stackrel{k^1}{k^2}}(t-t') \\
&& \label{contraction-II}
+
\sqrt{(k^1+1)(k^1+2)}
A^{(1+)}_{cm} A^{(1-)}_{\stackrel{m^1+1}{m^2}\stackrel{c^1+1}{c^2}} 
K_{\stackrel{k^1+1}{k^2},\stackrel{m^1}{m^2};\stackrel{m^1+1}{m^2},\stackrel{k^1+2}{k^2}}(t') 
K_{\stackrel{k^1+1}{k^2}\stackrel{c^1+1}{c^2};\stackrel{c^1}{c^2}\stackrel{k^1}{k^2}}(t-t')\\
&& \label{contraction-III}
+
\sqrt{(k^1+1)(k^1+2)}
A^{(1+)}_{\stackrel{c^1+1}{c^2}\stackrel{m^1+1}{m^2}} A^{(1-)}_{mc} 
K_{\stackrel{k^1+2}{k^2},\stackrel{m^1+1}{m^2};\stackrel{m^1}{m^2},\stackrel{k^1+1}{k^2}}(t') 
K_{\stackrel{k^1}{k^2}\stackrel{c^1}{c^2};\stackrel{c^1+1}{c^2}\stackrel{k^1+1}{k^2}}(t-t')\\
&& \label{contraction-IV}
+
(k^1+1)
A^{(1+)}_{\stackrel{c^1}{c^2+1}\stackrel{m^1}{m^2+1}} A^{(1-)}_{mc} 
K_{\stackrel{k^1+1}{k^2+1},\stackrel{m^1}{m^2+1};\stackrel{m^1}{m^2},\stackrel{k^1+1}{k^2}}(t') 
K_{\stackrel{k^1}{k^2}\stackrel{c^1}{c^2};\stackrel{c^1}{c^2+1},\stackrel{k^1}{k^2+1}}(t-t')\\
&& \label{contraction-V}
+
(k^1+1)
A^{(1+)}_{cm} A^{(1-)}_{\stackrel{m^1}{m^2+1}\stackrel{c}{c^1}{c^2+1}} 
K_{\stackrel{k^1+1}{k^2},\stackrel{m^1}{m^2};\stackrel{m^1}{m^2+1},\stackrel{k^1+1}{k^2+1}}(t') 
K_{\stackrel{k^1}{k^2+1},\stackrel{c^1}{c^2+1};\stackrel{c^1}{c^2},\stackrel{k^1}{k^2}}(t-t')
\Bigg\}
\\
\nonumber
&& + \mathcal O(\epsilon^0)\,.
\eea

The contribution from the choice $A^{(1-)}-A^{(1+)}$, first block has a similar form:
\bea
\nonumber
\Gamma_{1l,2.8}^\epsilon & = &
 - \frac{\theta}4 \int_\epsilon^\infty \frac{dt}{t} \int_0^t
dt' t' \, e^{-2t\sigma^2}  (1-\Omega^2)^2\\
&& \label{contraction-2I}
\hspace{-1.5cm}\times
\Bigg\{
(k^1+1)
A^{(1-)}_{cm} A^{(1+)}_{mc} 
K_{\stackrel{k^1}{k^2},\stackrel{m^1}{m^2};\stackrel{m^1}{m^2},\stackrel{k^1}{k^2}}(t') 
K_{\stackrel{k^1+1}{k^2}\stackrel{c^1}{c^2};\stackrel{c^1}{c^2}\stackrel{k^1+1}{k^2}}(t-t') \\
&& \label{contraction-2II}
\hspace{-.6cm} +
\sqrt{(k^1+1)(k^1+2)}
A^{(1-)}_{cm} A^{(1+)}_{\stackrel{m^1+1}{m^2}\stackrel{c^1+1}{c^2}} 
K_{\stackrel{k^1}{k^2},\stackrel{m^1}{m^2};\stackrel{m^1+1}{m^2},\stackrel{k^1+1}{k^2}}(t') 
K_{\stackrel{k^1+2}{k^2}\stackrel{c^1+1}{c^2};\stackrel{c^1}{c^2}\stackrel{k^1+1}{k^2}}(t-t')\\
&& \label{contraction-2III}
\hspace{-.6cm} +
\sqrt{(k^1+1)(k^1+2)}
A^{(1-)}_{\stackrel{c^1+1}{c^2}\stackrel{m^1+1}{m^2}} A^{(1+)}_{mc} 
K_{\stackrel{k^1+1}{k^2},\stackrel{m^1+1}{m^2};\stackrel{m^1}{m^2},\stackrel{k^1}{k^2}}(t') 
K_{\stackrel{k^1+1}{k^2}\stackrel{c^1}{c^2};\stackrel{c^1+1}{c^2}\stackrel{k^1+2}{k^2}}(t-t')\\
&& \label{contraction-2IV}
\hspace{-.6cm} +
(k^1+1)
A^{(1-)}_{\stackrel{c^1}{c^2+1}\stackrel{m^1}{m^2+1}} A^{(1+)}_{mc} 
K_{\stackrel{k^1}{k^2+1},\stackrel{m^1}{m^2+1};\stackrel{m^1}{m^2},\stackrel{k^1}{k^2}}(t') 
K_{\stackrel{k^1+1}{k^2}\stackrel{c^1}{c^2};\stackrel{c^1}{c^2+1},\stackrel{k^1+1}{k^2+1}}(t-t')\\
&& \label{contraction-2V}
\hspace{-.6cm} +
(k^1+1)
A^{(1-)}_{cm} A^{(1+)}_{\stackrel{m^1}{m^2+1}\stackrel{c^1}{c^2+1}} 
K_{\stackrel{k^1}{k^2},\stackrel{m^1}{m^2};\stackrel{m^1}{m^2+1},\stackrel{k^1}{k^2+1}}(t') 
K_{\stackrel{k^1+1}{k^2+1}\stackrel{c^1}{c^2+1};\stackrel{c^1}{c^2},\stackrel{k^1+1}{k^2}}(t-t')
\Bigg\}\,.
\eea
In order to calculate the contractions (\ref{contraction-I}-\ref{contraction-2V}), we distinguish between 
{\it leading} and {\it subleading} contributions. Leading contributions stem only from the leading terms of 
the infinite sums (\ref{app4}) and (\ref{app5}). In case of quadratic divergent contributions, the subleading
terms will be logarithmic divergent and need to be considered. The contractions (\ref{contraction-I}) and 
(\ref{contraction-2I}) allow for these subleading divergences.

Let us first consider the leading order contributions. We get the following results:

\begin{itemize}

\item {First contraction, term (\ref{contraction-I})}

\bea
\nonumber
\Gamma 
& = & - \frac{\theta}4 \int \frac{dt}{t} e^{-2t\sigma^2} \int dt' \, t' \, (1-\Omega^2)^2 
\Bigg\{ A^{(1+)}_{nk} A^{(1-)}_{kn} + A^{(2+)}_{nk} A^{(2-)}_{kn} \Bigg\}
\\
\nonumber
&&\times \frac1{t^2(1+\Omega^2)^2}\, \frac1{t(1+\Omega^2)} +\mathcal O(\epsilon^0)\\
\nonumber
& = & - \frac{\theta}8  \frac{(1-\Omega^2)^2}{(1+\Omega^2)^3}  
\int \frac{dt}{t^2} e^{-2t\sigma^2} \frac1{4\pi^2\theta^2} \int d^4x\,
A_\nu \star A^\nu\Big|_{(1+),(1-)} +\mathcal O(\epsilon^0)\\
\nonumber
& = & - \frac1{32\pi^2 \theta}  \frac{(1-\Omega^2)^2}{(1+\Omega^2)^3} \int d^4x\,
\left(\frac1{\epsilon} + \frac{\mu^2 \theta}2 \ln \epsilon
+ 4\Omega \ln\epsilon \right)\, A_\nu \star A^\nu \Big|_{(1+),(1-)}+\mathcal O(\epsilon^0)\\
& = & - \frac1{32\pi^2 \theta}  \rho^2 \int d^4x\,
\left(\frac1{\tilde \epsilon} + \frac{\tilde\mu^2}2 \ln \epsilon
+ \frac{4\Omega}{1+\Omega^2} \ln\epsilon \right)\, A_\nu \star A^\nu \Big|_{(1+),(1-)}
+\mathcal O(\epsilon^0)
\label{cI}
\eea


\item {Second contraction, term (\ref{contraction-II})}

\be
\Gamma =
\frac{\theta}{24} \rho^4 \ln\epsilon \sqrt{(m^1+1)(c^1+1)}
A^{(1+)}_{cm} A^{(1-)}_{\stackrel{m^1+1}{m^2}\stackrel{c^1+1}{c^2}} +\mathcal O(\epsilon^0)
\ee

\item {Contraction (\ref{contraction-III})}

\be
\Gamma_{1l,2.9}^\epsilon =
\frac{\theta}{24} \rho^4 \ln\epsilon \sqrt{(m^1+1)(c^1+1)} 
A^{(1+)}_{\stackrel{c^1+1}{c^2}\stackrel{m^1+1}{m^2}} A^{(1-)}_{mc} +\mathcal O(\epsilon^0)
\ee

\item {Contraction (\ref{contraction-IV})}

\be
\Gamma =
\frac{\theta}{48} \rho^4 \ln\epsilon \sqrt{(m^2+1)(c^2+1)}
A^{(1+)}_{\stackrel{c^1}{c^2+1}\stackrel{m^1}{m^2+1}} A^{(1-)}_{mc} +\mathcal O(\epsilon^0)
\ee

\item {Contraction (\ref{contraction-V})}

\be
\Gamma =
\frac{\theta}{48} \rho^4 \ln\epsilon \sqrt{(m^2+1)(c^2+1)}
A^{(1+)}_{cm} A^{(1-)}_{\stackrel{m^1}{m^2+1}\stackrel{c^1}{c^2+1}} +\mathcal O(\epsilon^0)
\ee

\end{itemize}
Since 
$$
\frac1{4\pi^2\theta^2} \int d^4 x \, A_\mu \star A^\mu = \sum_{m,c} \left( A^{(1+)}_{cm} A^{(1-)}_{mc}
+ A^{(2+)}_{cm} A^{(2-)}_{mc} \right),
$$
by $\Big|_{(1+),(1-)}$ we denote the restriction of expressions to the fields $A^{(1+)}$ and $A^{(1-)}$.
We have used this notation e.g. in Eq.~(\ref{cI}). The missing parts are due to the field content $A^{(2+)} - A^{(2-)}$
in (\ref{2nd}). They complement each other.

Sticking the above contributions together yields
\bea
\label{result-1}
\Gamma_{1l,2.9}^\epsilon & = & \frac{-1}{48\pi^2} \rho^2 
\left( \frac3{2\tilde \epsilon\, \theta} + \frac{3\tilde \mu^2}{4\theta} \ln\epsilon + \frac{ 6 \Omega}{1+\Omega^2}
\ln \epsilon \right) \int d^4 x\, A_\mu \star A^\mu\\
\nonumber
&&
+ \frac{\theta}{24} \rho^4 \ln\epsilon \sqrt{(m^1+1)(c^1+1)}
\left( A^{(1+)}_{cm} A^{(1-)}_{\stackrel{m^1+1}{m^2}\stackrel{c^1+1}{c^2}} + 
A^{(1+)}_{\stackrel{c^1+1}{c^2}\stackrel{m^1+1}{m^2}} A^{(1-)}_{mc} 
\right)\\
\nonumber
&&
+ \frac{\theta}{48}\rho^4 \ln\epsilon \sqrt{(m^2+1)(c^2+1)}
\left( A^{(1+)}_{cm} A^{(1-)}_{\stackrel{m^1}{m^2+1}\stackrel{c^1}{c^2+1}} + 
A^{(1+)}_{\stackrel{c^1}{c^2+1}\stackrel{m^1}{m^2+1}} A^{(1-)}_{mc} 
\right)\\
\nonumber
&& +\mathcal O(\epsilon^0)\,.
\eea
From contractions (\ref{contraction-2I})-(\ref{contraction-2V}) (i.e., field configuration $A^{(1-)}-A^{(1+)}$), 
we obtain the same result as in (\ref{result-1}) (i.e., field configuration $A^{(1+)}-A^{(1-)}$). The second block 
also gives the same contributions. Therefore, we obtain an overall factor of 4.

Next, we have to examine the subleading contributions. We have to start at the sum of Eqns.~(\ref{contraction-I}) 
and (\ref{contraction-2I}) where we want to extract the subleading divergences:
\bea
\nonumber
\Gamma_{1l,2.10}^\epsilon & = &  - \frac{\theta}4 \int_\epsilon^\infty \frac{dt}{t} \int_0^t
dt' t' \, e^{-2t\sigma^2}  (1-\Omega^2)^2 (k^1+1)\\
&& \nonumber
\times \Bigg\{
A^{(1+)}_{cm} A^{(1-)}_{mc} 
K_{\stackrel{k^1+1}{k^2},\stackrel{m^1}{m^2};\stackrel{m^1}{m^2},\stackrel{k^1+1}{k^2}}(t') 
K_{\stackrel{k^1}{k^2}\stackrel{c^1}{c^2};\stackrel{c^1}{c^2}\stackrel{k^1}{k^2}}(t-t') \\
\nonumber
&& + A^{(1-)}_{cm} A^{(1+)}_{mc} 
K_{\stackrel{k^1}{k^2},\stackrel{m^1}{m^2};\stackrel{m^1}{m^2},\stackrel{k^1}{k^2}}(t') 
K_{\stackrel{k^1+1}{k^2}\stackrel{c^1}{c^2};\stackrel{c^1}{c^2}\stackrel{k^1+1}{k^2}}(t-t') 
\Bigg\}\,.
\eea
Therefore, we expand the partial traces. For example, the second one yields:
\bea
\nonumber
&&\hspace{-1cm}
\sum_{k^1=0}^\infty (k^1+1) K_{k^1m^1;m^1k^1}(t') K_{k^1+1,c^1;c^1,k^1+1}(t-t') 
\sum_{k^2=0}^\infty  K_{k^2m^2;m^2k^2}(t') K_{k^2+1,c^2;c^2,k^2+1}(t-t') = \\
\nonumber
&& = X_\Omega(t')^{m^1+m^2+2} X_\Omega(t-t')^{c^1+c^2+3} e^{4 \Omega t}
\sum_{k^1,k^2} (k^1+1) \left( X_\Omega(t') X_\Omega(t-t') \right)^{k^1+k^2}\\
\nonumber
&& \hspace{.6cm} \times
\sum_{u=0}^{\min(m^1,k^1)} \binom{m^1}{u} \binom{k^1}{u}
e^{4\Omega t'u } \left( \frac{1-\Omega^2} {4\Omega}(1-e^{-4\Omega t'}) \right)^{2u}\\
\nonumber
&& \hspace{.6cm} \times
\sum_{v=0}^{\min(k^1+1,c^1)} \binom{k^1+1}{v} \binom{c^1}{v}
e^{4\Omega (t-t')v } \left( \frac{1-\Omega^2} {4\Omega} (1-e^{-4\Omega (t-t')})\right)^{2v}
\\
\nonumber
&& \hspace{.6cm} \times
\sum_{r=0}^{\min(m^2,k^2)} \binom{m^2}{r} \binom{k^2}{r}
e^{4\Omega t'r } \left( \frac{1-\Omega^2} {4\Omega}(1-e^{-4\Omega t'}) \right)^{2r}\\
\nonumber
&& \hspace{.6cm} \times
\sum_{s=0}^{\min(k^2,c^2)} \binom{k^2}{s} \binom{c^2}{s}
e^{4\Omega (t-t')s } \left( \frac{1-\Omega^2} {4\Omega} (1-e^{-4\Omega (t-t')})\right)^{2s}
\\
\nonumber
&& =
\left(
1 + 4 \Omega t - (m^1+m^2+2)(1+\Omega^2)t' - (c^1+c^2+3)(1+\Omega^2)(t-t')
\right)
\\
&& \hspace{.6cm}
\nonumber
\times \frac1{(1- X_\Omega(t') X_\Omega(t-t'))^3}\\
\nonumber
&& \hspace{.6cm}
+ (1-\Omega^2)^2 \bigg(
m^1 \frac{2t'^2}{t^4} + m^2 \frac{t'^2}{t^4} + c^1 \frac{2(t-t')^2}{t^4} + c^2 \frac{(t-t')^2}{t^4}
\bigg)
+\dots\\
\nonumber
&& =
\left(
1 + 4 \Omega t - (m^1+m^2+2)(1+\Omega^2)t' - (c^1+c^2+3)(1+\Omega^2)(t-t')
\right)
\\
&& \hspace{.6cm}
\nonumber
\times \left(
\frac1{(1+\Omega^2)^3 t^3} + \frac{3(1+\Omega^4)}{(1+\Omega^2)^4 t^2} - \frac{3t'(1-\Omega^2)^2}{(1+\Omega^2)^4t^3}
\right)
\\
\nonumber
&& \hspace{.6cm}
+ (1-\Omega^2)^2 \bigg(
m^1 \frac{2t'^2}{t^4} + m^2 \frac{t'^2}{t^4} + c^1 \frac{2(t-t')^2}{t^4} + c^2 \frac{(t-t')^2}{t^4}
\bigg)
+\dots
\eea
using $X_\Omega(t)^m = 1- (1+\Omega^2)m t + \mathcal O(t^2)$ and the geometric series given in Eqns.~(\ref{series}). 
Thus, we obtain for $\Gamma^\epsilon_{1l,2.10}$ to subleading order the following expressions:
\bea
\Gamma^\epsilon_{1l,2.10} & = & 
\nonumber
\frac{\theta}{12} \rho^4 \ln\epsilon 
\left( A^{(1+)}_{cm}A^{(1-)}_{mc} ( 2 c^1 + c^2 )
     + A^{(1-)}_{cm}A^{(1+)}_{mc} ( 2 c^1 + c^2 ) \right)\\
\nonumber
&& - \frac{\ln\epsilon}{24 \pi^2} \left(\frac{1-\Omega^2}{1+\Omega^2} \right)^2 \int d^4x\, 
\Bigg\{ -\frac{6\Omega}{\theta(1+\Omega^2)} A_\mu\star A^\mu
+ \frac9{4\theta} A_\mu\star A^\mu\\
\nonumber
&& \hspace{2cm}
+ \frac34 \tilde x^2(A_\mu\star A^\mu)\Bigg\}\Big|_{(1+),(1-)}\\
\nonumber
&&-\frac1{16\pi^2\theta} \ln\epsilon \left(\frac{1-\Omega^2}{1+\Omega^2} \right)^2 \int d^4 x \, 
A_\mu\star A^\mu\Big|_{(1+),(1-)} +\mathcal O(\epsilon^0)\\
& = & 
\label{subI}
-\frac{\ln\epsilon}{12\pi^2(1+\Omega^2)^2} \int d^4 x
\frac{(1-\Omega^2)^4}{(1+\Omega^2)^2} \left( -\frac14 \tilde x^2 (A_\mu \star A^\mu)
+\frac1{2\theta} A_\mu\star A^\mu \right) \Bigg|_{(1+),(1-)}\\  
\nonumber
&& \hspace{1.5cm}
+ \frac{\theta\ln\epsilon}{12 (1+\Omega^2)^2} \frac{(1-\Omega^2)^4}{(1+\Omega^2)^2}
A^{(1+)}_{cm} A^{(1-)}_{cm} ( c^1 + m^1)
\\
\label{subII}
&& - \frac{\ln\epsilon}{24 \pi^2} \left(\frac{1-\Omega^2}{1+\Omega^2} \right)^2 \int d^4x\, 
\Bigg\{ -\frac{6\Omega}{\theta(1+\Omega^2)} A_\mu\star A^\mu
+ \frac9{4\theta} A_\mu\star A^\mu\\
\nonumber
&& \hspace{2cm}
+ \frac34 \tilde x^2(A_\mu\star A^\mu)\Bigg\}\Big|_{(1+),(1-)}\\
\label{subIII}
&&-\frac1{16\pi^2\theta} \ln\epsilon \left(\frac{1-\Omega^2}{1+\Omega^2} \right)^2 \int d^4 x \, 
A_\mu\star A^\mu\Big|_{(1+),(1-)} +\mathcal O(\epsilon^0)
\eea

\paragraph{Summation of above contributions.}
Let us sum the contributions (\ref{1++}), (\ref{1--}), line 2 and 3 of (\ref{result-1}) and (\ref{subI})
taking into account the correct multiplicities. We obtain
\bea
\Gamma^\epsilon_{1l,2.11} & = &
-\frac{\ln\epsilon}{12\pi^2(1+\Omega^2)^2} \int d^4 x
    \frac{(1-\Omega^2)^4}{(1+\Omega^2)^2} \left( -\frac12 \tilde x^2 (A_\mu \star A^\mu)
    +\frac1{\theta} A_\mu\star A^\mu \right)\\  
\nonumber
&&+ \frac{\theta\ln\epsilon}{6(1+\Omega^2)^2} \frac{(1-\Omega^2)^4}{(1+\Omega^2)^2}
A^{(1+)}_{nk} A^{(1-)}_{kn} ( n^1 + k^1 )
\\
\nonumber
&&
+ \frac{\theta}{6} \left( \frac{1-\Omega^2}{1+\Omega^2} \right)^4 \ln\epsilon \sqrt{(m^1+1)(c^1+1)}
\left( A^{(1+)}_{cm} A^{(1-)}_{\stackrel{m^1+1}{m^2}\stackrel{c^1+1}{c^2}} + 
A^{(1+)}_{\stackrel{c^1+1}{c^2}\stackrel{m^1+1}{m^2}} A^{(1-)}_{mc} 
\right)\\
\nonumber
&&
+ \frac{\theta}{12} \left( \frac{1-\Omega^2}{1+\Omega^2} \right)^4 \ln\epsilon \sqrt{(m^2+1)(c^2+1)}
\left( A^{(1+)}_{cm} A^{(1-)}_{\stackrel{m^1}{m^2+1}\stackrel{c^1}{c^2+1}} + 
A^{(1+)}_{\stackrel{c^1}{c^2+1}\stackrel{m^1}{m^2+1}} A^{(1-)}_{mc} 
\right)\\
&&
-\frac{\theta}{12} \left( \frac{1-\Omega^2}{1+\Omega^2} \right)^4 \ln\epsilon \sqrt{m^1+1}\sqrt{c^1+1}\,
A^{(1+)}_{\stackrel{c^1}{c^2}\stackrel{m^1+1}{m^2}} A^{(1+)}_{\stackrel{m^1}{m^2}\stackrel{c^1+1}{c^2}}\\
\nonumber
&&
- \frac{\theta}{6} \left( \frac{1-\Omega^2}{1+\Omega^2} \right)^4 \ln\epsilon \,\sqrt{c^1}\sqrt{c^1+1} \,
A^{(1+)}_{\stackrel{c^1-1}{c^2}\stackrel{m^1}{m^2}} A^{(1+)}_{\stackrel{m^1}{m^2}\stackrel{c^1+1}{c^2}}\\
\nonumber
&&
- \frac{\theta}{12} \left( \frac{1-\Omega^2}{1+\Omega^2} \right)^4 \ln\epsilon \sqrt{m^1+1}\sqrt{c^1+1}\,
A^{(1-)}_{\stackrel{c^1+1}{c^2}\stackrel{m^1}{m^2}} A^{(1-)}_{\stackrel{m^1+1}{m^2}\stackrel{c^1}{c^2}}\\
\nonumber
&&
- \frac{\theta}{6} \left( \frac{1-\Omega^2}{1+\Omega^2} \right)^4 \ln\epsilon \,\sqrt{c^1}\sqrt{c^1+1} \,
A^{(1-)}_{\stackrel{c^1+1}{c^2}\stackrel{m^1}{m^2}} A^{(1-)}_{\stackrel{m^1}{m^2}\stackrel{c^1-1}{c^2}}\\
\nonumber
& = & -\frac{\ln\epsilon}{12\pi^2(1+\Omega^2)^2} \int d^4 x
    \frac{(1-\Omega^2)^4}{(1+\Omega^2)^2} \left( -\frac12 \tilde x^2 (A_\mu \star A^\mu)
    +\frac1{\theta} A_\mu\star A^\mu \right)\\  
\nonumber
&&+ \frac{\ln\epsilon}{12} \theta \left( \frac{1-\Omega^2}{1+\Omega^2} \right)^4
\Bigg\{
2 A^{(1+)}_{nk} A^{(1-)}_{kn} ( n^1 + k^1 + 1 ) - 2 A^{(1+)}_{nk} A^{(1-)}_{kn}\\
\nonumber
&& \hspace{2cm}
+ \sqrt{(m^1+1)(c^1+1)}
\left( A^{(1+)}_{cm} A^{(1-)}_{\stackrel{m^1+1}{m^2}\stackrel{c^1+1}{c^2}}
+ A^{(1+)}_{\stackrel{c^1+1}{c^2}\stackrel{m^1+1}{m^2}} A^{(1-)}_{mc} 
\right)\\
&&\hspace{2cm}
- \sqrt{m^1+1}\sqrt{c^1+1}\, \left(
A^{(1+)}_{\stackrel{c^1}{c^2}\stackrel{m^1+1}{m^2}} A^{(1+)}_{\stackrel{m^1}{m^2}\stackrel{c^1+1}{c^2}}
+
A^{(1-)}_{\stackrel{c^1+1}{c^2}\stackrel{m^1}{m^2}} A^{(1-)}_{\stackrel{m^1+1}{m^2}\stackrel{c^1}{c^2}}
\right)\\
\nonumber
&&\hspace{2cm}
- 2 \sqrt{c^1}\sqrt{c^1+1} \, \left(
A^{(1+)}_{\stackrel{c^1-1}{c^2}\stackrel{m^1}{m^2}} A^{(1+)}_{\stackrel{m^1}{m^2}\stackrel{c^1+1}{c^2}}
+
A^{(1-)}_{\stackrel{c^1+1}{c^2}\stackrel{m^1}{m^2}} A^{(1-)}_{\stackrel{m^1}{m^2}\stackrel{c^1-1}{c^2}}
\right)
\Bigg\}\\
\nonumber
&&+ \frac{\theta}{12} \left( \frac{1-\Omega^2}{1+\Omega^2} \right)^4 \ln\epsilon 
\Bigg\{
\sqrt{(m^2+1)(c^2+1)}
\left(A^{(1+)}_{\stackrel{c^1}{c^2+1}\stackrel{m^1}{m^2+1}} A^{(1-)}_{mc}
+A^{(1+)}_{cm} A^{(1-)}_{\stackrel{m^1}{m^2+1}\stackrel{c^1}{c^2+1}}\right)\\
\nonumber
&& \hspace{2cm}
+\sqrt{(m^1+1)(c^1+1)}
\left( A^{(1+)}_{cm} A^{(1-)}_{\stackrel{m^1+1}{m^2}\stackrel{c^1+1}{c^2}} + 
A^{(1+)}_{\stackrel{c^1+1}{c^2}\stackrel{m^1+1}{m^2}} A^{(1-)}_{mc} \right)
\Bigg\}
\\
\nonumber
& = & -\frac{\ln\epsilon}{12\pi^2(1+\Omega^2)^2} \int d^4 x
    \frac{(1-\Omega^2)^4}{(1+\Omega^2)^2} \left( -\frac12 \tilde x^2 (A_\mu \star A^\mu)
    +\frac3{2\theta} A_\mu\star A^\mu \right)\Bigg|_{(1+),(1-)}
\\  
\nonumber
&&- \frac{\ln\epsilon}{12\pi^2 } \left( \frac{1-\Omega^2}{1+\Omega^2} \right)^4
\int d^4 x\, \Bigg\{
- \frac14 \{\tilde x_\nu, A^\nu\}_\star \star \{\tilde x_\mu, A^\mu\}_\star
\\
\nonumber
&& \hspace{1cm}
- \frac12 \tilde x_\nu \star \tilde x_\mu \star A^\nu \star A^\mu 
\Bigg\} \Bigg|_{(1+),(1-)}
\\
\nonumber
&& 
+ \frac{\ln\epsilon}{12} \left( \frac{1-\Omega^2}{1+\Omega^2} \right)^4 
\Bigg\{
\frac1{2\pi^2} \int d^4 x\, (\tilde x^2 A^\sigma)\star A_\sigma\\
\nonumber
&& \hspace{1cm}
- 2 \theta\, A^{(1+)}_{nk} A^{(1-)}_{kn} ( n^1 + n^2 + 1) 
\Bigg\}\Bigg|_{(1+),(1-)}
\\
\label{result-3}
& = & - \frac{\ln\epsilon}{12\pi^2 } \left( \frac{1-\Omega^2}{1+\Omega^2} \right)^4
\int d^4 x\, \Bigg\{
-\frac12 \tilde x^2 (A_\mu \star A^\mu) + \frac3{2\theta} A_\mu\star A^\mu \\
\nonumber
&&
- \frac14 \{\tilde x_\nu, A^\nu\}_\star \star \{\tilde x_\mu, A^\mu\}_\star
- \frac12 \tilde x_\nu \star \tilde x_\mu \star A^\nu \star A^\mu
- \frac14 \tilde x_\nu \star A^\mu \star \tilde x^\nu \star A_\mu
\Bigg\} \Bigg|_{(1+),(1-)}\,.
\eea
We have made use of the matrix base expressions quoted in the Appendix.

\subsubsection{Second order result}

In the end, we can add up all the different terms. The result to second order is found to be
\bea
\Gamma_{1l,2}^\epsilon 
& = & \frac{- \rho^2 \ln\epsilon }{12 \pi^2} \int d^4x\, 
\Bigg\{
\bigg(-\frac38 ( \tilde X_\mu \star \tilde X^\mu \star \tilde X_\nu \star \tilde X^\nu - (\tilde x^2)^2)
\\
\nonumber
&& \hspace{1.5cm}
+ \frac34 \tilde x ^2(A_\mu\star A^\mu) + \frac32 \tilde x^2 (\tilde xA)
\bigg)\\
&&
+\bigg( 
 \frac1{\tilde \epsilon} \frac{3}{2 \theta} A_\mu \star A^\mu
+ \frac{\tilde \mu^2}{\theta} \ln\epsilon \,\frac{3}{4} A_\mu \star A^\mu
\\
\nonumber
&&
\hspace{1.5cm}+
\ln\epsilon\, \frac{6 \Omega}{\theta (1+\Omega^2)} A_\mu \star A^\mu
- \frac14 \rho^2  A_\mu \star \tilde x_\nu \star A^\mu \star \tilde x^\nu
\bigg)
\\
&&
+ \bigg(
- \frac{6\Omega}{\theta(1+\Omega^2)}  A_\mu \star A^\mu 
+ \frac9{4\theta} A_\mu \star A^\mu\\
\nonumber
&& \hspace{1.5cm} +
\frac34 \tilde x^2(A_\mu \star A^\mu)
-\frac3{2\theta} A_\mu\star A^\mu\\
&& \hspace{1.5cm}
-\frac12 \rho^2 \tilde x^2 (A_\mu \star A^\mu) 
- \frac34 \rho^2 A_\mu\star A^\mu
\nonumber
\\
\nonumber
&& \hspace{1.5cm} -
\rho^2 \left(
 \frac14 \{\tilde x_\nu, A^\nu\}_\star \star \{\tilde x_\mu, A^\mu\}_\star 
+ \frac12 \tilde x_\mu \tilde x_\nu (A^\mu\star A^\nu)
\right)
\bigg)
\\
&&
+ \bigg(
\frac34  A_\mu \star A^\mu\star \{\tilde x_\nu, \star A^\nu\}_\star 
+ \frac34 \{\tilde x_\nu , A^\nu\}_\star \star \{\tilde x_\mu , A^\mu \}_\star \bigg)
\Bigg\}\\
\nonumber
&& + \mathcal O(\epsilon^0)\,.
\eea


\subsection{Third order}

In third order, the one-loop effective action is given by
\bea
\Gamma^\epsilon_{1l,3} & = & \frac{\theta^3}{16} 
\int_\epsilon^\infty \frac{dt}{t} \int_0^t dt' \int_0^{t'} dt'' \, t'' e^{-2t\sigma^2}
\sum_{k,l,m,n,a,b,c,d,g,f,u,v} 
\delta_{n+a,m+b} \delta_{d+g,c+f} \delta_{v+l,u+k}\\
\nonumber
&& \times
V_{kl;mn} K(t'')_{nm;ab} V_{ba;cd} K(t'-t'')_{dc;gf} V_{fg;uv} K(t-t')_{vu;lk}  \,.
\eea
There are two different divergent contributions.


\paragraph{$\mathbf{A-A-\tilde X^2}$.}

In order to obtain a divergent contribution both $A$ fields have to be taken from the same black, one with index "-" and one 
with index "+". 
So, let us consider the case $A^{(1+)}A^{(1-)}B^2$, where both, $A^{(1-)}$ and $A^{(1+)}$ are taken from the first
block of Eq.~(\ref{sppc}). Then, we obtain
\bea
\Gamma^\epsilon_{1l,3.1} & = & \frac{\theta^3}{16} 
\int_\epsilon^\infty \frac{dt}{t} \int_0^t dt' \int_0^{t'} dt'' \, t'' e^{-2t\sigma^2}
(1+\Omega^2)(1-\Omega^2)^2 \frac2{\theta} \delta_{n+a,m+b} \delta_{d+g,c+f} \delta_{v+l,u+k}\\
\nonumber
&& \times
\sqrt{n^1}A_{lm}^{(1+)} \delta_{\stackrel{k^1}{k^2}\stackrel{n^1-1}{n^2}}
\sqrt{d^1+1} A_{ac}^{(1-)} \delta_{\stackrel{d^1+1}{d^2}\stackrel{b^1}{b^2}}
K_{nm;ab}(t'')K_{dc;gf}(t'-t'')K_{vu;lk}(t-t')\\
\nonumber
&& \times
\bigg(
(B_\mu\star B^\mu-\tilde x^2)_{gu}\delta_{vf}+ (B_\mu\star B^\mu-\tilde x^2)_{vf}\delta_{gu}
\bigg) 
\\
& = & \frac{\theta^2}8 
\int_\epsilon^\infty \frac{dt}{t} \int_0^t dt' \int_0^{t'} dt'' \, t'' e^{-2t\sigma^2}
(1+\Omega^2)(1-\Omega^2)^2 
\\
\nonumber
&& \times
\sum_{c,d,l,m} (d^1+1) A^{(1+)}_{lm} A^{(1-)}_{mc}(B_\nu\star B^\nu -\tilde x^2)_{cl}
\\
\nonumber
&& \hspace{1.5cm}
\times 
K_{\stackrel{d^1+1}{d^2},\stackrel{m^1}{m^2};\stackrel{m^1}{m^2},\stackrel{d^1+1}{d^2}}(t'')
K_{dc;cd}(t'-t'')   K_{dl;ld}(t-t') + \mathcal O(\epsilon^0)\\
& = & \frac{\theta^2}8 
\int_\epsilon^\infty \frac{dt}{t} \int_0^t dt' \int_0^{t'} dt'' \,  t'' e^{-2t\sigma^2}
(1+\Omega^2)(1-\Omega^2)^2 
\frac1{t^2(1+\Omega^2)^2} \frac1{t(1+\Omega^2)}\\
\nonumber
&& \times
\sum_{c,l,m} A^{(1+)}_{lm} A^{(1-)}_{mc} (B_\nu\star B^\nu -\tilde x^2)_{cl}
+\mathcal O(\epsilon^0)\\
& = &  \frac{-\ln\epsilon}{12\pi^2}\frac1{16}\int d^4 x \left( \frac{1-\Omega^2}{1+\Omega^2} \right)^2
A^{(1+)}\star A^{(1-)} \star (B_\nu\star B^\nu -\tilde x^2) + \mathcal O (\epsilon^0)\,.
\eea
There is an equal contribution coming from the second block. Therefore, we have a multiplicity factor of $12$, since
we can additionally rearrange the fields in the product "$AAB^2$". Thus, we get
\be
\Gamma^\epsilon_{1l,3.1} = \frac{-\ln\epsilon}{16\pi^2} \int d^4 x \rho^2
A_\mu\star A^\mu \star (A_\nu\star A^\nu + 2\tilde x A) + \mathcal O(\epsilon^0)\,.
\ee

\paragraph{$\mathbf{A-A-A}$.}

All the fields have to be chosen from the same block of Eq.~(\ref{sppc}). Otherwise the contributions 
are finite. Either all three fields are from the same oscillator or only two of them. In the latter case the
signs belonging to the same oscillator have to be saturated.

We first examine the expression related to the choice $A^{(1+)}-A^{(1-)}-A^{(1-)}$, where all the fields are 
taken from the first block. The calculation yields
\bea
\Gamma^\epsilon_{1l,3.2} & = & \frac{\theta^3}{16} \int_\epsilon^\infty \frac{dt}{t} \int_0^t dt' \int_0^{t'}dt''t'' e^{-2\sigma^2t}
\delta_{n+a,m+b} \delta_{d+p,c+q} \delta_{s+l,r+k}\\
\nonumber
&& \times
(1-\Omega^2)^3(-i) \left(\frac2{\theta}\right)^{3/2} \sqrt{n^1}A_{lm}^{(1+)} \delta_{\stackrel{k^1}{k^2}\stackrel{n^1-1}{n^2}}
\sqrt{d^1+1} A_{ac}^{(1-)} \delta_{\stackrel{d^1+1}{d^2}\stackrel{b^1}{b^2}}\\
\nonumber
&& \times
\sqrt{s^1+1} A^{(1-)}_{pr} \delta_{\stackrel{s^1+1}{s^2}\stackrel{q^1}{q^2}}
K(t'')_{nm;ab}K(t'-t'')_{dc;pq}K(t-t')_{sr;lk}\\
\nonumber
& = & -i\left(\frac2{\theta}\right)^{3/2} \frac{\theta^3}{16}(1-\Omega^2)^3
\int_\epsilon^\infty \frac{dt}{t} \int_0^t dt' \int_0^{t'}dt''t'' e^{-2\sigma^2t}
\sum_{m,p,r} A_{rm}^{(1+)} A_{mp}^{(1-)} A^{(1-)}_{\stackrel{p^1+1}{p^2}\stackrel{r^1}{r^2}} \\
\nonumber
&& \times 
\sum_{n^1} (n^1+1)^{3/2} K(t'')_{n^1+1,m^1;m^1,n^1+1} K(t'-t'')_{n^1p^1;p^1+1,n^1+1} K(t-t')_{n^1r^1;r^1n^1} \\
\nonumber
&& \times 
\sum_{n^2} K(t'')_{n^2m^2;m^2n^2}K(t'-t'')_{n^2p^2;p^2n^2} K(t-t')_{n^2r^2;r^2n^2} + \mathcal O(\epsilon^0)\\
\label{aaa1}
& = & i \ln\epsilon \sqrt{ \frac2{\theta} } \, \rho^4 \frac{\theta^2}{96}
\sum_{m,p,r} \sqrt{p^1+1} A_{rm}^{(1+)} A_{mp}^{(1-)} A^{(1-)}_{\stackrel{p^1+1}{p^2}\stackrel{r^1}{r^2}} 
+ \mathcal O(\epsilon^0)
\,.
\eea
The expression for $A^{(1+)}A^{(1-)}A^{(1+)}$ is of a slightly different form:
\bea
\label{aaa2}
\Gamma^\epsilon_{1l,3.3} = - i \ln\epsilon \sqrt{ \frac2{\theta} } \, \frac{\theta^2(1-\Omega^2)^4}{4(1+\Omega^2)^4} \frac1{24}
\sum_{m,p,r} \sqrt{r^1+1} A_{rm}^{(1+)} A_{mp}^{(1-)} A^{(1+)}_{\stackrel{p^1}{p^2}\stackrel{r^1+1}{r^2}}
+ \mathcal O(\epsilon^0)\,.
\eea
Both terms, (\ref{aaa1}) and (\ref{aaa2}) appear 6 times. Note the difference in the overall sign.
Therefore, we obtain the contribution
\bea
\nonumber
\Gamma^\epsilon_{1l,3.4} & = & \Gamma^\epsilon_{1l,3.2} + \Gamma^\epsilon_{1l,3.3}  \\
& = & 
i\ln\epsilon \sqrt{\frac2{\theta}} \frac{\theta^2}{16} \left(\frac{1-\Omega^2}{1+\Omega^2}\right)^4
\\
\nonumber
&& \times \sum_{rmp} \sqrt{p^1+1}\bigg(
A^{(1+)}_{rm} A^{(1-)}_{mp} A^{(1-)}_{\stackrel{p^1+1}{p^2}\stackrel{r^1}{r^2}} -
A^{(1+)}_{pm} A^{(1-)}_{mr} A^{(1+)}_{\stackrel{r^1}{r^2}\stackrel{p^1+1}{p^2}}
\bigg) + \mathcal O(\epsilon^0)\,.
\eea
Also the field content of the form $A^{(1+)}A^{(1-)}A^{(2\pm)}$ produces a divergent contribution, e.g.,
\bea
\label{aaa3}
\Gamma^\epsilon_{1l,3.4} =  i \ln\epsilon \sqrt{ \frac2{\theta} } \, \frac{\theta^2(1-\Omega^2)^4}{4(1+\Omega^2)^4} \frac1{24}
\sum_{m,p,r} \sqrt{l^2+1} A_{\stackrel{l^1}{l^2+1}\stackrel{m^1}{m^2}}^{(1+)} A_{mc}^{(1-)} 
A^{(2-)}_{cl}
+ \mathcal O(\epsilon^0)\,.
\eea

Comparing (\ref{matrix1}) and (\ref{matrix2}), we see that the above expressions (plus the ones we
have omitted here) are equal to 
\bea
\Gamma^\epsilon_{1l,3.5} = \frac{\ln\epsilon}{12 \pi^2}\frac38  \left(\frac{1-\Omega^2}{1+\Omega^2}\right)^4
\int d^4 x\, \bigg(
A_\mu\star A^\mu \star \{\tilde x_\nu , A^\nu\}_\star - \tilde x_\nu \star A_\mu \star A^\nu \star A^\mu 
\bigg) + \mathcal O (\epsilon^0)\,.
\eea
\vspace{.6cm}

Summing up all the partial contributions, we obtain for the action in third order:
\bea
\Gamma^\epsilon_{1l,3} & = & 
\frac{-\ln\epsilon}{12\pi^2} \int d^4x\, 
\Bigg\{
\frac34 \rho^2 \bigg( (A_\mu\star A^\mu)^{\star 2}
+ A_\mu \star A^\mu\star \{ \tilde x_\nu, \star A^\nu\}_\star \bigg)\\
\nonumber
&& \hspace{.8cm}
- \frac12 \rho^4 \bigg( 
\tilde x_\nu \star A_\mu \star A^\nu\star A^\mu
+ A_\mu\star A^\mu \star \{\tilde x_\nu, \star A^\nu\}_\star \bigg)
\Bigg\} + \mathcal O(\epsilon^0)\,.
\eea


\subsection{Fourth order}

The fourth order expression of the effective action reads:
\bea
\Gamma^\epsilon_{1l,4} & = & - 
\frac{\theta^4}{32} \int_\epsilon^\infty \frac{dt}{t} e^{-2\sigma^2 t}\,\int_0^t dt' \int_0^{t'} dt'' 
    \int_0^{t''} dt''' \, t''' e^{-2\sigma^2 t} \\
\nonumber
&& \times \sum \delta_{n+a,m+b} \delta_{d+e,c+f} \delta_{h+i,g+j} \delta_{q+l,p+k}
V_{kl;mn} K(t''')_{nm;ab} V_{ba;cd}\\
\nonumber
&& \hspace{.5cm}
\times  K(t''-t''')_{dc,ef} V_{fe;gh} K(t'-t'')_{hg;ij} V_{ji;pq} K(t-t')_{qp;lk}\,.
\eea
There is only one divergent contribution stemming from the field content $A-A-A-A$. All the fields have to come from
the same block. Fields from the second oscillator may mix with fields from the first in a single expression, 
but the signs need to be saturated for each oscillator.

The explicit calculation yields the result:
\bea
\Gamma^\epsilon_{1l,4} 
& = & \frac{-\ln\epsilon}{12\pi^2} \int d^4x\, \frac1{8} \rho^4
\left( - 2(A_\mu \star A^\mu)^{\star 2} - A_\mu \star A_\nu \star A^\mu \star A^\nu \right)\\
\nonumber
&& + \mathcal O(\epsilon^0)\,.
\eea


\subsection{\label{res}Summed up result}

Summing up the order-by-order result, we end up at the final expression for the gauge field action:

\bea
\label{result2}
\Gamma_{1l}^\epsilon & = & \frac{1}{192\pi^2}  \int d^4x\, \Bigg\{
\frac{24}{\tilde \epsilon \, \theta} (1-\rho^2)(\tilde X_\nu \star \tilde X^\nu -\tilde x^2)\\
\nonumber
&&
+ \ln\epsilon \bigg(
\frac{12}{\theta} (1-\rho^2) (\tilde \mu^2-\rho^2)(\tilde X_\nu \star \tilde X^\nu -\tilde x^2) 
\\
\nonumber
&& \hspace{1.3cm}
+ 6(1-\rho^2)^2 \big( (\tilde X_\mu\star \tilde X^\mu)^{\star 2}-(\tilde x^2)^2 \big)
- \rho^4  F_{\mu\nu} F^{\mu\nu}
\bigg)
\Bigg\}\,,
\eea
where the field strength is given by
\be
F_{\mu\nu} = -i[\tilde x_\mu, A_\nu]_\star +i [\tilde x_\nu, A_\mu]_\star - i [A_\mu, A_\nu]_\star
\,.
\ee

\section{\label{conclusions}Conclusions}

Our main result is summarised in Eqn.~(\ref{result2}): Both, the linear in  $\epsilon$ and the logarithmic
in $\epsilon$ divergent term turn out to be gauge invariant. The logarithmically divergent part is an interesting
candidate for a renormalisable gauge interaction. We note that the resulting action has been proposed 
by R.~Wulkenhaar and one of us (H.G.) in previous reports. 
As far as we know, this action did not appear before in string theory.
The sign of the term quadratic in the covariant coordinates may change depending on whether $\tilde\mu^2\lessgtr \rho^2$.
This reflects a phase transition. In a forthcoming work (H.G. and H.~Steinacker, in preparation), we were able to
analyse in detail an action like (\ref{result2}) in two dimensions. The case $\Omega=1$ ($\rho=0$) is of course of 
particular interest. One obtains a matrix model. We shall return to a study of these models 
in a forthcoming publication \cite{Grosse:2007xx}. 
In the limit $\Omega\to 0$, we obtain just the standard deformed Yang-Mills action.
Furthermore, the action ~(\ref{result2}) allows to study the limit $\theta\to \infty$.

In addition, we will attempt to study the perturbative quantisation.
One of the problems of quantising action (\ref{result2}) is connected to the tadpole contribution,
which is non-vanishing and hard to eliminate. The Paris group arrived at similar conclusions.


\subsection*{Acknowledgements}

We want to express our gratitude to R.~Wulkenhaar for valuable contributions at an early stage of this work. 
H.G. thanks him for a long standing collaboration.
One of us (H.G.) enjoyed discussions with the Paris group (V.~Rivasseau et al.) in 
Oberwolfach.


\begin{appendix}
\section*{Appendix}

\renewcommand{\theequation}{A-\arabic{equation}}
\setcounter{equation}{0}

Useful geometric series and variants:
\bea
\nonumber
\sum_{n=0}^\infty X^n & = & \frac1{1-X},\\
\nonumber
\sum_{n=0}^\infty n X^n & = & \frac{X}{(1-X)^2},\\
\label{series}
\sum_{n=0}^\infty (n+1) X^n & = & \frac1{(1-X)^2},\\
\nonumber
\sum_{n=0}^\infty (n+1)^2 X^n & = & \frac{1+X}{(1-X)^3},\\
\nonumber
\sum_{n=0}^\infty (n+1) n X^n & = & \frac{2X}{(1-X)^3}.
\eea

Derivatives yield in the matrix basis
\bea
(\partial_\nu \psi)(x) & = & 
\sum_{p,q \in \mathbb{N}^2} \psi_{pq} \partial_\nu f_{pq}(x) 
\nonumber
\\
&=& \sum_{p,q \in \mathbb{N}^2} \psi_{pq} \Big(
\frac{\partial a^1}{\partial x^\nu} 
\frac{\partial f_{pq}}{\partial a^1}
+ \frac{\partial \bar{a}^1}{\partial x^\nu} 
\frac{\partial f_{pq}}{\partial \bar{a}^1}
+ \frac{\partial a^2}{\partial x^\nu} 
\frac{\partial f_{pq}}{\partial a^2}
+ \frac{\partial \bar{a}^2}{\partial x^\nu} 
\frac{\partial f_{pq}}{\partial \bar{a}^2}\Big)
\nonumber
\\
&= &\frac{1}{\sqrt{2}\theta} \sum_{p,q \in \mathbb{N}^2} 
\Big(
(\delta_{\nu,1}+\mathrm{i}\delta_{\nu,2})
(f_{pq} \star \bar{a}^1 - \bar{a}^1 \star f_{pq})
+ (\delta_{\nu,1}-\mathrm{i}\delta_{\nu,2})
(a^1 \star f_{pq} - f_{pq} \star a^1)
\nonumber
\\
&& +(\delta_{\nu,3}+\mathrm{i}\delta_{\nu,4})
(f_{pq} \star \bar{a}^2 - \bar{a}^2 \star f_{pq})
+ (\delta_{\nu,3}-\mathrm{i}\delta_{\nu,4})
(a^2 \star f_{pq} - f_{pq} \star a^2)\Big)\psi_{pq} 
\nonumber
\\
&=& \frac{1}{\sqrt{2\theta}} \sum_{p,q \in \mathbb{N}^2} 
\Big(
(\delta_{\nu,1}+\mathrm{i}\delta_{\nu,2})
(\sqrt{q^1} f_{p^1,q^1-1} - \sqrt{p^1+1} f_{p^1+1,q^1}) f_{p^2q^2}
\nonumber
\\[-2ex]
&&\hspace*{6em} + (\delta_{\nu,1}-\mathrm{i}\delta_{\nu,2})
(\sqrt{p^1} f_{p^1-1,q^1} - \sqrt{q^1+1}f_{p^1,q^1+1}) f_{p^2q^2}
\nonumber
\\
&& \hspace*{6em}  +(\delta_{\nu,3}+\mathrm{i}\delta_{\nu,4})
f_{p^1q^1} (\sqrt{q^2} f_{p^2 q^2-1}  - \sqrt{p^2+1} f_{p^2+1,q^2})
\nonumber
\\
&& \hspace*{6em} + (\delta_{\nu,3}-\mathrm{i}\delta_{\nu,4})
f_{p^1q^1}(\sqrt{p^2} f_{p^2-1,q^2} 
- \sqrt{q^2+1}f_{p^2,q^2+1})\Big)\psi_{pq} 
\nonumber
\\
&=& \frac{1}{\sqrt{2\theta}} \sum_{p,q \in \mathbb{N}^2} 
\Big(
(\delta_{\nu,1}+\mathrm{i}\delta_{\nu,2})
(\sqrt{q^1+1} \psi_{\stackrel{p^1}{p^2}\stackrel{q^1+1}{q^2}} 
- \sqrt{p^1} \psi_{\stackrel{p^1-1}{p^2}\stackrel{q^1}{q^2}} )
\nonumber
\\[-2ex]
&&\hspace*{6em} + (\delta_{\nu,1}-\mathrm{i}\delta_{\nu,2})
(\sqrt{p^1+1 }\psi_{\stackrel{p^1+1}{p^2}\stackrel{q^1}{q^2}} 
 - \sqrt{q^1}\psi_{\stackrel{p^1}{p^2}\stackrel{q^1-1}{q^2}} )
\nonumber
\\
&& \hspace*{6em}  +(\delta_{\nu,3}+\mathrm{i}\delta_{\nu,4})
(\sqrt{q^2+1} \psi_{\stackrel{p^1}{p^2}\stackrel{q^1}{q^2+1}} 
- \sqrt{p^2} \psi_{\stackrel{p^1}{p^2-1}\stackrel{q^1}{q^2}} )
\nonumber
\\
\label{app1}
&& \hspace*{6em} + (\delta_{\nu,3}-\mathrm{i}\delta_{\nu,4})
(\sqrt{p^2+1} \psi_{\stackrel{p^1}{p^2+1}\stackrel{q^1}{q^2}} 
- \sqrt{q^2} \psi_{\stackrel{p^1}{p^2}\stackrel{q^1}{q^2-1}} )
\Big)f_{pq} \;.
\eea

We also compute $\tilde x_\mu \cdot \psi$ in the matrix basis
\bea
&&2 \tilde x_\nu \cdot \psi(x) = 
\nonumber
\\
&=&\sum_{p,q\in \mathbb{N}^2} \psi_{pq} \Big(
\delta_{\nu,1}(\theta^{-1})_{12} (x^2 \star f_{pq} + f_{pq} \star x^2)(x) 
+ \delta_{\nu,2}(\theta^{-1})_{21} (x^1 \star f_{pq} + f_{pq} \star
x^1)(x) 
\nonumber
\\*[-1ex]
&&\qquad \qquad
+\delta_{\nu,3}(\theta^{-1})_{34} (x^4 \star f_{pq} + f_{pq} \star x^4)(x) 
+ \delta_{\nu,4}(\theta^{-1})_{43} (x^3 \star f_{pq} + f_{pq} \star
x^3)(x) \Big)
\nonumber
\\
&=&\frac{\mathrm{i}}{\sqrt{2}\theta} 
\sum_{p,q\in \mathbb{N}^2} \psi_{pq} \Big(
- (\delta_{\nu,1}+\mathrm{i}\delta_{\nu,2})
(\bar{a}^1\star f_{pq} + f_{pq} \star \bar{a}^1)
+ (\delta_{\nu,1}-\mathrm{i}\delta_{\nu,2})
(a^1 \star f_{pq} + f_{pq} \star a^1)
\nonumber
\\*[-1ex]
&&\qquad \qquad
- (\delta_{\nu,3}+\mathrm{i}\delta_{\nu,4})
(\bar{a}^2\star f_{pq} + f_{pq} \star \bar{a}^2)
+ (\delta_{\nu,3}-\mathrm{i}\delta_{\nu,4})
(a^2 \star f_{pq} + f_{pq} \star a^2)\Big)(x)
\nonumber
\\
&=&\frac{\mathrm{i}}{\sqrt{2\theta}} 
\sum_{p,q\in \mathbb{N}^2} \psi_{pq} \Big(
- (\delta_{\nu,1}+\mathrm{i}\delta_{\nu,2})
(\sqrt{p^1+1} f_{\stackrel{p^1+1}{p^2}\stackrel{q^1}{q^2}} 
+ \sqrt{q^1} 
f_{\stackrel{p^1}{p^2}\stackrel{q^1-1}{q^2}} )
\nonumber
\\*[-1.5ex]
&&\qquad\qquad\qquad 
+ (\delta_{\nu,1}-\mathrm{i}\delta_{\nu,2})
(\sqrt{p^1} f_{\stackrel{p^1-1}{p^2}\stackrel{q^1}{q^2}} 
+ \sqrt{q^1+1}f_{\stackrel{p^1}{p^2}\stackrel{q^1+1}{q^2}})
\nonumber
\\
&&\qquad\qquad\qquad
- (\delta_{\nu,3}+\mathrm{i}\delta_{\nu,4})
(\sqrt{p^2+1} f_{\stackrel{p^1}{p^2+1}\stackrel{q^1}{q^2}} 
+ \sqrt{q^2} f_{\stackrel{p^1}{p^2}\stackrel{q^1}{q^2-1}} )
\nonumber
\\
\nonumber
&&\qquad\qquad\qquad
+ (\delta_{\nu,3}-\mathrm{i}\delta_{\nu,4})
(\sqrt{p^2} f_{\stackrel{p^1}{p^2-1}\stackrel{q^1}{q^2}} 
+ \sqrt{q^2+1}f_{\stackrel{p^1}{p^2}\stackrel{q^1}{q^2+1}})\Big)(x)
\\
&=&\frac{\mathrm{i}}{\sqrt{2\theta}} 
\sum_{p,q\in \mathbb{N}^2} \Big(
- (\delta_{\nu,1}+\mathrm{i}\delta_{\nu,2})
(\sqrt{p^1} \psi_{\stackrel{p^1-1}{p^2}\stackrel{q^1}{q^2}} 
+ \sqrt{q^1+1} \psi_{\stackrel{p^1}{p^2}\stackrel{q^1+1}{q^2}} )
\nonumber
\\*[-1.5ex]
&&\qquad\qquad\qquad 
+ (\delta_{\nu,1}-\mathrm{i}\delta_{\nu,2})
(\sqrt{p^1+1} \psi_{\stackrel{p^1+1}{p^2}\stackrel{q^1}{q^2}} 
+ \sqrt{q^1} \psi_{\stackrel{p^1}{p^2}\stackrel{q^1-1}{q^2}})
\nonumber
\\
&&\qquad\qquad\qquad
- (\delta_{\nu,3}+\mathrm{i}\delta_{\nu,4})
(\sqrt{p^2} \psi_{\stackrel{p^1}{p^2-1}\stackrel{q^1}{q^2}} 
+ \sqrt{q^2+1} \psi_{\stackrel{p^1}{p^2}\stackrel{q^1}{q^2+1}} )
\nonumber
\\
\label{app2}
&&\qquad\qquad\qquad
+ (\delta_{\nu,3}-\mathrm{i}\delta_{\nu,4})
(\sqrt{p^2+1} \psi_{\stackrel{p^1}{p^2+1}\stackrel{q^1}{q^2}} 
+ \sqrt{q^2} \psi_{\stackrel{p^1}{p^2}\stackrel{q^1}{q^2-1}})\Big) f_{pq}(x)\;.
\eea

\subsection*{Partial traces with two kernels}

We do not imply the double index notation here.
\bea
\nonumber
&&\sum_{n=0}^\infty K_{nm;mn}(t') K_{n+1,c;c,n+1}(t-t') = \\
\nonumber
&&=\sum_n \sum_{u=0}^{\min(m,n)} \binom{m}{u} \binom{n}{u}
e^{2\Omega t' ( 1 + 2u) } (1-e^{-4\Omega t'})^{2u}  \left( \frac{1-\Omega^2} {4\Omega} \right)^{2u}
X_\Omega(t')^{n+m+1} \\
\nonumber
&& \hspace{.6cm}
\times \sum_{v=0}^{\min(n+1,c)} \binom{n+1}{v} \binom{c}{v}
e^{2\Omega (t-t') ( 1 + 2v) } (1-e^{-4\Omega (t-t')})^{2v}  \left( \frac{1-\Omega^2} {4\Omega} \right)^{2v}
X_\Omega(t-t')^{n+c+2}\\
\nonumber
&& = X_\Omega(t')^{m+1} X_\Omega(t-t')^{c+2}
\sum_n \left( X_\Omega(t') X_\Omega(t-t') \right)^n\\
\nonumber
&& \hspace{.6cm} \times
\sum_{u=0}^{\min(m,n)} \binom{m}{u} \binom{n}{u}
e^{2\Omega t' ( 1 + 2u) } (1-e^{-4\Omega t'})^{2u}  \left( \frac{1-\Omega^2} {4\Omega} \right)^{2u}\\
\nonumber
&& \hspace{.6cm} \times
\sum_{v=0}^{\min(n+1,c)} \binom{n+1}{v} \binom{c}{v}
e^{2\Omega (t-t') ( 1 + 2v) } (1-e^{-4\Omega (t-t')})^{2v}  \left( \frac{1-\Omega^2} {4\Omega} \right)^{2v}
\\
& = & \frac1{t(1+\Omega^2)} + \mathcal O(t^0,t'^0)
\,,
\label{app4}
\eea
and we also need to consider the following partial sum:
\bea
\nonumber
&&\sum_{n=0}^\infty \sqrt{n+1} K_{nm;m+1,n+1}(t') K_{n+1,c;c,n+1}(t-t') =\\
\nonumber
&& = \sum_n \sqrt{n+1}
\sum_{u=0}^{\min(n,m)} \sqrt{ \binom{n+1}{u+1} \binom{n}{u} \binom{m+1}{u+1}\binom{m}{u} }
e^{2\Omega t'(2+2u)}\\
\nonumber
&& \hspace{1.5cm} \times
 (1-e^{-4\Omega t'})^{2u+1} \left( \frac{1-\Omega^2}{4\Omega} \right)^{2u+1}
X_\Omega(t')^{n+m+2}\\
\nonumber
&& \hspace{.6cm}
\times \sum_{v=0}^{\min(n+1,c)} \binom{n+1}{v} \binom{c}{v}
e^{2\Omega (t-t') ( 1 + 2v) } (1-e^{-4\Omega (t-t')})^{2v}  \left( \frac{1-\Omega^2} {4\Omega} \right)^{2v}
X_\Omega(t-t')^{n+c+2}\\
\nonumber
&& = X_\Omega(t')^{m+2} X_\Omega(t-t')^{c+2} \sqrt{m+1} \sum_n \left( X_\Omega(t') X_\Omega(t-t') \right)^n\\
\nonumber
&& \hspace{.6cm} \times
\sum_{u=0}^{\min(n,m)}  \binom{n+1}{u+1} \binom{m}{u} 
e^{2\Omega t'(2+2u)}(1-e^{-4\Omega t'})^{2u+1} \left( \frac{1-\Omega^2}{4\Omega} \right)^{2u+1}\\
\nonumber
&& \hspace{.6cm}
\times \sum_{v=0}^{\min(n+1,c)} \binom{n+1}{v} \binom{c}{v}
e^{2\Omega (t-t') ( 1 + 2v) } (1-e^{-4\Omega (t-t')})^{2v}  \left( \frac{1-\Omega^2} {4\Omega} \right)^{2v}
\\
&& =
\sqrt{m+1} \frac{1-\Omega^2}{(1+\Omega^2)^2} \frac{t'}{t^2} + \mathcal O(t'^0, t^0)\,.
\label{app5}
\eea


\subsection*{Expressions in the matrix basis}

\bea
\label{matrix2-4}
\frac1{4\pi^2\theta^2} \int d^4 x \, 4\,  \tilde x_\mu \star \tilde x^\mu \star A_\nu \star A^\nu =
\sum_{p,q\in \mathbb{N}^2} \frac8{\theta}(A_\mu\star A^\mu)_{\stackrel{q^1}{q^2}\stackrel{q^1}{q^2}}
(q^1+q^2+1)
\eea


\bea
\nonumber
&& \hspace{-2cm}
\frac1{4\pi^2\theta^2} \int d^4x \, \{\tilde x_\nu, A^\nu\}_\star \star \{\tilde x_\mu, A^\mu\}_\star =
\\
\label{matrix2-1}
&& \hspace{-1cm} = 
\frac1{\theta} \Bigg(
A^{(1+)}_{\stackrel{p^1}{p^2}\stackrel{q^1}{q^2}}
   	A^{(1-)}_{\stackrel{q^1}{q^2}\stackrel{p^1}{p^2}}
	\left( p^1 + q^1 + 1
	\right)
+A^{(2+)}_{\stackrel{p^1}{p^2}\stackrel{q^1}{q^2}}
   	A^{(2-)}_{\stackrel{q^1}{q^2}\stackrel{p^1}{p^2}}
	\left( p^2 + q^2 + 1
	\right)\\
&& \nonumber
+ \sqrt{p^1} \sqrt{q^1} A^{(1+)}_{\stackrel{p^1-1}{p^2}\stackrel{q^1-1}{q^2}}	A^{(1-)}_{qp}
+ \sqrt{p^1} \sqrt{q^1} A^{(1+)}_{pq} A^{(1-)}_{\stackrel{q^1-1}{q^2}\stackrel{p^1-1}{p^2}}	
\\
&& \nonumber
+ \sqrt{p^2} \sqrt{q^2} A^{(2+)}_{\stackrel{p^1}{p^2-1}\stackrel{q^1}{q^2-1}}	A^{(2-)}_{qp}
+ \sqrt{p^2} \sqrt{q^2} A^{(2+)}_{pq} A^{(2-)}_{\stackrel{q^1}{q^2-1}\stackrel{p^1}{p^2-1}}\\
&& \nonumber
- \sqrt{p^1} \sqrt{q^1} A^{(1+)}_{\stackrel{p^1-1}{p^2}\stackrel{q^1}{q^2}}	A^{(1+)}_{\stackrel{q^1-1}{q^2}\stackrel{p^1}{p^2}}
- \sqrt{p^1} \sqrt{p^1+1} A^{(1+)}_{\stackrel{p^1-1}{p^2}\stackrel{q^1}{q^2}}	A^{(1+)}_{\stackrel{q^1}{q^2}\stackrel{p^1+1}{p^2}}
\\
&& \nonumber
- \sqrt{p^2} \sqrt{q^2} A^{(2+)}_{\stackrel{p^1}{p^2-1}\stackrel{q^1}{q^2}}	A^{(2+)}_{\stackrel{q^1}{q^2-1}\stackrel{p^1}{p^2}}
- \sqrt{p^2} \sqrt{p^2+1} A^{(2+)}_{\stackrel{p^1}{p^2-1}\stackrel{q^1}{q^2}}	A^{(2+)}_{\stackrel{q^1}{q^2}\stackrel{p^1}{p^2+1}}
\\
&&\nonumber
- \sqrt{p^1} \sqrt{q^1} A^{(1-)}_{\stackrel{p^1}{p^2}\stackrel{q^1-1}{q^2}}	A^{(1-)}_{\stackrel{q^1}{q^2}\stackrel{p^1-1}{p^2}}
- \sqrt{p^1} \sqrt{p^1+1} A^{(1-)}_{\stackrel{p^1+1}{p^2}\stackrel{q^1}{q^2}}	A^{(1-)}_{\stackrel{q^1}{q^2}\stackrel{p^1-1}{p^2}}
\\
&&\nonumber
- \sqrt{p^2} \sqrt{q^2} A^{(2-)}_{\stackrel{p^1}{p^2}\stackrel{q^1}{q^2-1}}	A^{(2-)}_{\stackrel{q^1}{q^2}\stackrel{p^1}{p^2-1}}
- \sqrt{p^2} \sqrt{p^2+1} A^{(2-)}_{\stackrel{p^1}{p^2+1}\stackrel{q^1}{q^2}}	A^{(2-)}_{\stackrel{q^1}{q^2}\stackrel{p^1}{p^2-1}}
\\
&&\nonumber
- \sqrt{p^1} \sqrt{q^2} A^{(1+)}_{\stackrel{p^1-1}{p^2}\stackrel{q^1}{q^2}}	A^{(2+)}_{\stackrel{q^1}{q^2-1}\stackrel{p^1}{p^2}}
- \sqrt{p^1} \sqrt{p^2} A^{(1+)}_{\stackrel{p^1-1}{p^2-1}\stackrel{q^1}{q^2}}	A^{(2+)}_{\stackrel{q^1}{q^2}\stackrel{p^1}{p^2}}
\\
&&\nonumber
- \sqrt{q^1} \sqrt{p^2} A^{(1+)}_{\stackrel{p^1}{p^2-1}\stackrel{q^1}{q^2}}	A^{(2+)}_{\stackrel{q^1-1}{q^2}\stackrel{p^1}{p^2}}
- \sqrt{q^1} \sqrt{q^2} A^{(1+)}_{pq}	A^{(2+)}_{\stackrel{q^1-1}{q^2-1}\stackrel{p^1}{p^2}}
\\
&&\nonumber
+ \sqrt{p^1} \sqrt{q^2} A^{(1+)}_{\stackrel{p^1-1}{p^2}\stackrel{q^1}{q^2-1}}	A^{(2-)}_{qp}
+ \sqrt{p^1} \sqrt{p^2} A^{(1+)}_{\stackrel{p^1-1}{p^2}\stackrel{q^1}{q^2}}	A^{(2-)}_{\stackrel{q^1}{q^2}\stackrel{p^1}{p^2-1}}
\\
&&\nonumber
+ \sqrt{q^1} \sqrt{q^2} A^{(1+)}_{\stackrel{p^1}{p^2}\stackrel{q^1}{q^2-1}}	A^{(2-)}_{\stackrel{q^1-1}{q^2}\stackrel{p^1}{p^2}}
+ \sqrt{q^1} \sqrt{p^2} A^{(1+)}_{pq}	A^{(2-)}_{\stackrel{q^1-1}{q^2}\stackrel{p^1}{p^2-1}}
\\
&&\nonumber
+ \sqrt{q^1} \sqrt{p^2} A^{(1-)}_{\stackrel{p^1}{p^2-1}\stackrel{q^1-1}{q^2}}	A^{(2+)}_{qp}
+ \sqrt{q^1} \sqrt{q^2} A^{(1-)}_{\stackrel{p^1}{p^2}\stackrel{q^1-1}{q^2}}	A^{(2+)}_{\stackrel{q^1}{q^2-1}\stackrel{p^1}{p^2}}
\\
&&\nonumber
+ \sqrt{p^1} \sqrt{p^2} A^{(1-)}_{\stackrel{p^1}{p^2-1}\stackrel{q^1}{q^2}}	A^{(2+)}_{\stackrel{q^1}{q^2}\stackrel{p^1-1}{p^2}}
+ \sqrt{p^1} \sqrt{q^2} A^{(1-)}_{pq}	A^{(2+)}_{\stackrel{q^1}{q^2-1}\stackrel{p^1-1}{p^2}}
\\
&&\nonumber
- \sqrt{p^1} \sqrt{p^2} A^{(1-)}_{pq}	A^{(2-)}_{\stackrel{q^1}{q^2}\stackrel{p^1-1}{p^2-1}}
- \sqrt{p^1} \sqrt{q^2} A^{(1-)}_{\stackrel{p^1}{p^2}\stackrel{q^1}{q^2-1}}	A^{(2-)}_{\stackrel{q^1}{q^2}\stackrel{p^1-1}{p^2}}
\\
&&\nonumber
- \sqrt{q^1} \sqrt{p^2} A^{(1-)}_{\stackrel{p^1}{p^2}\stackrel{q^1-1}{q^2}}	A^{(2-)}_{\stackrel{q^1}{q^2}\stackrel{p^1}{p^2-1}}
- \sqrt{q^1} \sqrt{q^2} A^{(1-)}_{\stackrel{p^1}{p^2}{q^1-1}{q^2-1}}	A^{(2-)}_{qp}
\Bigg)
\eea


\bea
&&
\nonumber
\frac1{4\pi^2\theta^2} \int d^4 x\, 4\,  \tilde x_\nu \star \tilde x_\mu \star A^\nu\star A^\mu =
\\
\label{matrix2-2}
&& = \frac2{\theta} \Bigg( 
A^{(1+)}_{qp}A^{(1-)}_{pq}(q^1+p^1+1) + A^{(2+)}_{qp}A^{(2-)}_{pq}(q^2+p^2+1) \\
&& \nonumber \hspace{.5cm}
+ \sqrt{q^1 q^2} (A^{(2-)}\star A^{(1+)})_{\stackrel{q^1-1}{q^2}\stackrel{q^1}{q^2-1}}
- \sqrt{q^1 q^2} (A^{(2+)}\star A^{(1+)})_{\stackrel{q^1-1}{q^2-1}\stackrel{q^1}{q^2}}\\
&& \nonumber\hspace{.5cm}
+ \sqrt{q^1 q^2} (A^{(1+)}\star A^{(2-)})_{\stackrel{q^1-1}{q^2}\stackrel{q^1}{q^2-1}}
- \sqrt{q^1 q^2} (A^{(1+)}\star A^{(2+)})_{\stackrel{q^1-1}{q^2-1}\stackrel{q^1}{q^2}}\\
&& \nonumber\hspace{.5cm}
+ \sqrt{q^1 q^2} (A^{(2+)}\star A^{(1-)})_{\stackrel{q^1}{q^2-1}\stackrel{q^1-1}{q^2}}
- \sqrt{q^1 q^2} (A^{(2-)}\star A^{(1-)})_{\stackrel{q^1}{q^2}\stackrel{q^1-1}{q^2-1}}\\
&& \nonumber\hspace{.5cm}
+ \sqrt{q^1 q^2} (A^{(1-)}\star A^{(2+)})_{\stackrel{q^1}{q^2-1}\stackrel{q^1-1}{q^2}}
- \sqrt{q^1 q^2} (A^{(1-)}\star A^{(2-)})_{\stackrel{q^1}{q^2}\stackrel{q^1-1}{q^2-1}}\\
&& \nonumber\hspace{.5cm}
- \sqrt{q^1(q^1+1)} (A^{(1-)}\star A^{(1-)})_{\stackrel{q^1+1}{q^2}\stackrel{q^1-1}{q^2}}
- \sqrt{q^1(q^1+1)} (A^{(1+)}\star A^{(1+)})_{\stackrel{q^1-1}{q^2}\stackrel{q^1+1}{q^2}}\\
&& \nonumber\hspace{.5cm}
- \sqrt{q^2(q^2+1)} (A^{(2-)}\star A^{(2-)})_{\stackrel{q^1}{q^2+1}\stackrel{q^1}{q^2-1}}
- \sqrt{q^2(q^2+1)} (A^{(2+)}\star A^{(2+)})_{\stackrel{q^1}{q^2-1}\stackrel{q^1}{q^2+1}}
\Bigg)
\eea


\bea
&&
\nonumber
\frac1{4\pi^2\theta^2} \int d^4x\, 4 (\tilde x^2 \cdot A^\sigma)\star A_\sigma =
\\
\label{matrix2-3}
&& = \frac2{\theta} \Bigg\{ 
2 A^\sigma_{pq} (A_\sigma)_{qp} (p^1+p^2+1)\\
\nonumber 
&& \hspace{1cm}
+ \sqrt{p^1+1} \sqrt{q^1+1} \left(
A^{(1+)}_{\stackrel{p^1+1}{p^2}\stackrel{q^1+1}{q^2}} A^{(1-)}_{qp} + A^{(1+)}_{pq} A^{(1-)}_{\stackrel{q^1+1}{q^2}\stackrel{p^1+1}{p^2}}
\right)\\
\nonumber
&& \hspace{1cm}
+ \sqrt{p^1+1} \sqrt{q^1+1} \left(
A^{(2+)}_{\stackrel{p^1+1}{p^2}\stackrel{q^1+1}{q^2}} A^{(2-)}_{qp} + A^{(2+)}_{pq} A^{(2-)}_{\stackrel{q^1+1}{q^2}\stackrel{p^1+1}{p^2}}
\right)\\
\nonumber
&& \hspace{1cm}
+ \sqrt{p^2+1} \sqrt{q^2+1} \left(
A^{(1+)}_{\stackrel{p^1}{p^2+1}\stackrel{q^1}{q^2+1}} A^{(1-)}_{qp} + A^{(1+)}_{pq} A^{(1-)}_{\stackrel{q^1}{q^2+1}\stackrel{p^1}{p^2+1}}
\right)\\
\nonumber 
&& \hspace{1cm}
+ \sqrt{p^2+1} \sqrt{q^2+1} \left(
A^{(2+)}_{\stackrel{p^1}{p^2+1}\stackrel{q^1}{q^2+1}} A^{(2-)}_{qp} + A^{(2+)}_{pq} A^{(2-)}_{\stackrel{q^1}{q^2+1}\stackrel{p^1}{p^2+1}}
\right)
\Bigg\}
\eea


\bea
\nonumber
&&
\frac1{4\pi^2\theta^2} \int d^4x \, \tilde x_\nu \star A_\mu \star A^\nu \star A^\mu = 
\\
\nonumber
&&
=
\frac{i}{\sqrt{2\theta}} \sum_q \Bigg\{\\
\nonumber
&& \hspace{.5cm}
\sqrt{q^1+1} \bigg(
- (A_\mu\star A^{(1+)} \star A^\mu)_{\stackrel{q^1}{q^2}\stackrel{q^1+1}{q^2}}
+ (A_\mu\star A^{(1-)} \star A^\mu)_{\stackrel{q^1+1}{q^2}\stackrel{q^1}{q^2}}
\bigg)\\
\label{matrix1}
&& \hspace{.4cm}
+\sqrt{q^2+1} \bigg(
- (A_\mu\star A^{(2+)} \star A^\mu)_{\stackrel{q^1}{q^2}\stackrel{q^1}{q^2+1}}
+ (A_\mu\star A^{(2-)} \star A^\mu)_{\stackrel{q^1}{q^2+1}\stackrel{q^1}{q^2}}
\bigg)
\Bigg\}
,
\eea
where e.g.
\bea
(A_\mu\star A^{(1+)} \star A^\mu)_{\stackrel{q^1}{q^2}\stackrel{q^1+1}{q^2}} & = &
\frac12 \bigg( 
A^{(1+)}_{qa} A^{(1+)}_{ab} A^{(1-)}_{\stackrel{b^1}{b^2}\stackrel{q^1+1}{q^2}} +
A^{(1-)}_{qa} A^{(1+)}_{ab} A^{(1+)}_{\stackrel{b^1}{b^2}\stackrel{q^1+1}{q^2}} \\
\nonumber
&&
+
A^{(2+)}_{qa} A^{(1+)}_{ab} A^{(2-)}_{\stackrel{b^1}{b^2}\stackrel{q^1+1}{q^2}} +
A^{(2-)}_{qa} A^{(1+)}_{ab} A^{(2+)}_{\stackrel{b^1}{b^2}\stackrel{q^1+1}{q^2}}
\bigg)\,.
\eea


\bea 
\nonumber
&&
\frac1{4\pi^2\theta^2}
\int d^4x  (\tilde x_\nu \star A^\nu + A^\nu \star \tilde x_\nu) \star A_\mu \star A^\mu =
\\
\nonumber
&&
=
\frac{i}{\sqrt{2\theta}} \Bigg\{
- \sqrt{p^1} A^{(1+)}_{\stackrel{p^1-1}{p^2}\stackrel{q^1}{q^2}} (A_\mu\star A^\mu)_{qp}
- \sqrt{q^1+1} A^{(1+)}_{\stackrel{p^1}{p^2}\stackrel{q^1+1}{q^2}} (A_\mu\star A^\mu)_{qp}\\
\nonumber
&& \hspace{.6cm}
+ \sqrt{p^1+1} A^{(1-)}_{\stackrel{p^1+1}{p^2}\stackrel{q^1}{q^2}} (A_\mu\star A^\mu)_{qp}
+ \sqrt{q^1} A^{(1-)}_{\stackrel{p^1}{p^2}\stackrel{q^1-1}{q^2}} (A_\mu\star A^\mu)_{qp}
\\
\nonumber
&& \hspace{.6cm}
- \sqrt{p^2} A^{(2+)}_{\stackrel{p^1}{p^2-1}\stackrel{q^1}{q^2}} (A_\mu\star A^\mu)_{qp}
- \sqrt{q^2+1} A^{(2+)}_{\stackrel{p^1}{p^2}\stackrel{q^1}{q^2+1}} (A_\mu\star A^\mu)_{qp}\\
\label{matrix2}
&& \hspace{.6cm}
+ \sqrt{p^2+1} A^{(2-)}_{\stackrel{p^1}{p^2+1}\stackrel{q^1}{q^2}} (A_\mu\star A^\mu)_{qp}
+ \sqrt{q^2} A^{(2-)}_{\stackrel{p^1}{p^2}\stackrel{q^1}{q^2-1}} (A_\mu\star A^\mu)_{qp}
\Bigg\}
\eea

\end{appendix}


\bibliographystyle{../latex-styles/utphys}
\bibliography{../tuw}

\end{document}